\documentclass[10pt,journal,compsoc]{IEEEtran}
\IEEEoverridecommandlockouts
\usepackage{cite}
\usepackage{color}
\usepackage{amsmath,amssymb,amsfonts}
\usepackage{amsthm}
\DeclareMathOperator*{\Median}{median}
\usepackage{empheq}
\usepackage[ulem={normalem,normalbf}]{changes}
\setuptodonotes{fancyline}
\newtheorem{definition}{Definition}
\newtheorem{theorem}{Theorem}

\newenvironment*{prooff}{\noindent \textbf{Proof}.}{\hfill $\blacksquare$ \vskip 4mm}
\usepackage{stfloats}
\usepackage[ruled,vlined]{algorithm2e}
\SetKwInput{KwInit}{Initialization}
\usepackage{graphicx}
\usepackage{caption}
\usepackage{subcaption}
\usepackage{textcomp}
\usepackage{xcolor}
\usepackage{mathtools}
\usepackage{array}
\usepackage{bm}
\usepackage{booktabs}
\def\BibTeX{{\rm B\kern-.05em{\sc i\kern-.025em b}\kern-.08em
		T\kern-.1667em\lower.7ex\hbox{E}\kern-.125emX}}

\usepackage{hyperref}
\usepackage{xurl}
\usepackage{enumitem}
\usepackage{tikz}
\usepackage{pgfplots}
\pgfplotsset{compat=1.18}

\begin{document}

\title{Adaptive Personalized Federated Reinforcement Learning for RIS-Assisted Aerial Relays in SAGINs with Fluid Antennas}

\author{Yuxuan Yang, Bin Lyu, \textit{Senior Member,~IEEE}, and Abbas Jamalipour, \textit{Fellow,~IEEE}
\thanks{Yuxuan Yang and Abbas Jamalipour are with the School of Electrical and Computer Engineering, University of Sydney, Sydney,
NSW, 2006 Australia (e-mail: yuxuan.yang@sydney.edu.au; a.jamalipour@ieee.org).}
\thanks{Bin Lyu is with the School of Communications and Information Engineering, Nanjing University of Posts and Telecommunications, Nanjing, 210003, China (e-mail: blyu@njupt.edu.cn).}
}

\IEEEtitleabstractindextext{
\begin{abstract}
    Space–air–ground integrated networks (SAGINs) interconnect satellites, uncrewed aerial vehicles (UAVs), and ground devices to enable flexible and ubiquitous wireless services. The integration of reconfigurable intelligent surfaces (RISs) and fluid antenna systems (FASs) further enhances radio environment controllability. However, the tight integration of cross-layer facilities and radio enhancement technologies leads to pronounced environmental dynamics and heterogeneity, posing fundamental challenges for system modeling and optimization in large-scale SAGINs. This paper investigates a SAGIN in which low Earth orbit (LEO) satellite constellations communicate with multiple ground hotspots via RIS-assisted UAV relays, serving both FAS-equipped and conventional users. A system model is developed that explicitly captures satellite mobility, UAV trajectories, RIS phase control, and heterogeneous user reception capabilities. Accordingly, a multi-hotspot downlink rate maximization problem is studied, whose solvability is analyzed through a hierarchical Stackelberg game. To address heterogeneous and time-varying multi-hotspot environments, an adaptive personalized federated reinforcement learning (FRL) algorithm is proposed for adaptive optimization of UAV trajectories and RIS phase controls. Simulation results demonstrate superior performance and validate the effectiveness of personalization in dynamic heterogeneous SAGIN scenarios.
   
\end{abstract}
}

\maketitle
\section{Introduction}\label{intro}
\IEEEPARstart{S}{pace}–air–ground integrated networks (SAGINs) have attracted significant attention as a fundamental framework for 6G wireless communications \cite{10745905}. By interconnecting heterogeneous network infrastructures deployed at different altitude layers, SAGINs substantially enhance the ubiquitous connectivity, worldwide access, and integrated communication–computing capabilities \cite{8869705}. 

As implied by the name, SAGINs typically comprise three layers. In the space layer, satellites, particularly LEO satellites, play a critical role, with ultra-dense LEO constellations emerging as a promising networking paradigm \cite{9982369}. For example, OneWeb plans to expand its constellation from 648 to about 7,000 satellites, while SpaceX’s Starlink has already deployed over 2,000 satellites \cite{10228912}. In the air layer, high-altitude platforms (HAPs) and uncrewed aerial vehicles (UAVs) are widely employed to assist communications, serving as aerial base stations (BSs), edge servers, and relays \cite{8918497}, owing to reliable line-of-sight (LoS) links and high mobility.
In the ground layer, user devices exhibit diverse reception characteristics and activity patterns, and the quality of service they experience often serves as a primary indicator of the system performance.

SAGINs have been widely studied from various perspectives. Yao et al. investigated a secure maritime SAGIN, where beamforming and UAV trajectory are jointly optimized under imperfect channel state information (CSI) \cite{11123630}. Fan et al. developed a SAGIN-enabled framework for data collection by jointly optimizing satellite association, HAP and UAV trajectories, and UAV transmit power\cite{11077742}. In \cite{10679223}, task hosting, offloading decisions, association control, and computing resource allocation were jointly optimized to achieve cost-efficient computation offloading in SAGIN. In \cite{10820863}, data collection in SAGIN was studied, where UAV trajectories and network configurations are jointly optimized to minimize the average system age of information.

To enhance radio-environment controllability and mitigate severe propagation loss and blockage in SAGINs, reconfigurable intelligent surfaces (RISs) have been widely adopted to enable passive signal reflection with programmable phase control \cite{9424177}. Accordingly, extensive studies have investigated RIS-assisted SAGINs. In \cite{10771971}, a RIS-assisted multi-hop UAV relay network was studied, where UAV trajectories and RIS phase shifts were jointly optimized for throughput maximization. Nguyen et al. \cite{9822386} investigated a RIS–UAV relay-assisted transmission framework for SAGINs with adaptive link switching to combat cloud blockage and atmospheric turbulence. Lukito et al. \cite{10702554} considered a satellite IoT system integrated with a UAV-mounted STAR-RIS, and jointly optimized UAV trajectory, phase shifts, and power allocation using a Dinkelbach-based approach. In \cite{10848151}, RIS-aided downlink MIMO transmission in ultra-dense LEO satellite networks was studied, where user association, beamforming, and RIS phase shifts were alternately optimized.

Moreover, an emerging antenna technology, known as the fluid antenna system (FAS), has been introduced to enhance signal reception at the user end. First proposed in \cite{9264694}, FAS enables flexible position (port) adjustment within a predefined region, thereby improving antenna gain, radiation patterns, and link quality, and is regarded as a key enabling technology for 6G communications \cite{11247926}. Ghadi et al. analyzed RIS-reflected channels received by FAS-equipped users, focusing on the spatial correlation across antenna positions \cite{10539238}. Wang et al. explored visions of AI-empowered FASs, especially employing deep reinforcement learning (DRL) to address high-dimensional challenges\cite{10599127}. Despite the  complementarity between RIS and FAS, the integration of RIS and FAS in the context of SAGINs remains in an evolving stage, which motivates a systematic investigation of their combination.

\begin{figure*}[ht]
		\centering
		\includegraphics[scale=0.46]{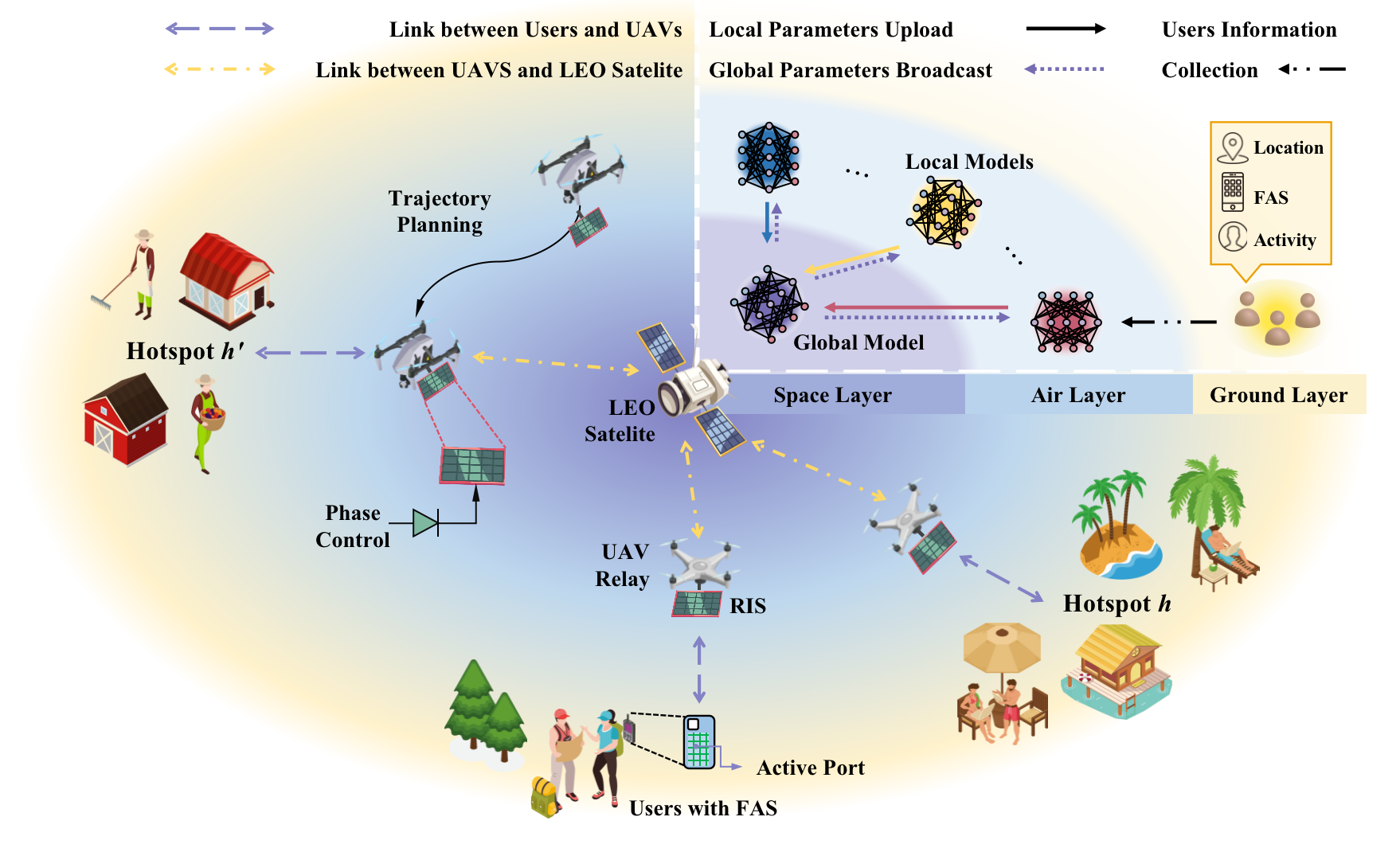}
		\renewcommand{\figurename}{Fig.}
		\caption{Hierarchical SAGIN architecture with FRL framework.}
		\label{ssm}
	\end{figure*}

In parallel with advances in physical-layer technologies, including RIS and FAS, SAGINs have increasingly embraced AI-empowered solutions to cope with the pronounced dynamics and randomness introduced by heterogeneous entities across different layers. \cite{11123630,11077742} utilized successive convex approximation (SCA) based approaches, which are not suitable for instantaneous random CSI. \cite{10679223,10820863,10771971} applied DRL solutions. However, for multi-agent scenarios such as multi-UAV trajectory optimization \cite{10820863} and multi-RIS phase control \cite{10771971}, DRL-based methods adopt centralized training, leading to substantial communication overhead and secure risks.

Unlike DRL, which follows a data-to-model paradigm, federated reinforcement learning (FRL) adopts a model-to-data framework where agents train locally and only exchange model parameters with a global server \cite{pmlr-v162-khodadadian22a}. This paradigm naturally fits the hierarchical structure of SAGINs, where satellites act as global servers and UAVs or HAPs serve as local agents supporting ground hotspots. By transmitting only model information rather than raw data, FRL improves the reliability of satellite links. Motivated by this, Liu et al. studied computation offloading in SAGINs, and proposed a FRL framework to minimize long-term energy consumption\cite{9941490}. In \cite{10989513}, time-sensitive traffic management in SAGINs was investigated, where traffic scheduling and load balancing are jointly optimized with a FRL solution. Qin et al. proposed a differentiated FRL framework, achieving improved throughput, delay, and packet loss performance in SAGINs\cite{10502326}.

Existing studies indicates that the combined use of RIS and FAS can effectively alleviate performance bottlenecks in SAGINs by strengthening satellite links, exploiting UAV mobility, and enhancing user-side reception. However, despite extensive studies on satellites, UAVs, RISs, and FASs individually, their integration within a unified SAGIN framework remains underexplored. Investigating such an integrated SAGIN is therefore necessary to fully exploit their benefits in support of ubiquitous 6G connectivity. Moreover, environmental heterogeneity is often oversimplified, and the channel heterogeneity introduced by FAS is rarely studied. This gap is particularly critical for FRL-based control, as a single global policy cannot be optimal across heterogeneous environments \cite{Killian_Konidaris_Doshi-Velez_2017}.

Motivated by the analysis above, this paper investigates a SAGIN where a LEO satellite constellation communications with multiple ground hotspots via RIS-assisted UAV relays, as shown in Fig. \ref{ssm}. In particular, users in each hotspot are randomly activated, with a subset equipped with FAS. Under an FRL framework, each UAV serves as a local agent, training its local policy model. The satellite constellation acts as a global server to aggregate local models and update a global policy. Each UAV further employs an adaptive personalization mechanism to integrate local and global policies for joint optimization of UAV trajectories and RIS phase controls. FAS users perform port activation based on received signals, and all users feed back location, FAS capability, and activation status to their associated UAVs for experience generation. The contributions of this study are summarized as follows:
    
\begin{itemize}

    \item We develop a unified SAGIN model that integrates LEO satellites, RIS-assisted UAV relays, and ground users with and without FASs, characterizing channel heterogeneity across multiple hotspots. A system-level downlink rate maximization problem is thereby formulated under UAV mobility and RIS phase control constraints.

    \item A hierarchical Stackelberg game is formulated to capture the intrinsic leader–follower interactions among satellites, UAV relays, and ground users, through which the solvability of the formulated optimization problem is theoretically established.

    \item A personalized FRL framework is proposed to address environmental heterogeneity, where adaptive personalization is achieved without introducing additional network structures, enabling efficient joint optimization of UAV trajectories and RIS phase controls.

    \item Extensive simulations are conducted to characterize the role of personalization in FRL under heterogeneous SAGIN environments, revealing how adaptive integration of local and global policies influences learning stability and system performance. The results provide practical insights into personalized FRL mechanisms for large-scale, heterogeneous SAGINs.

\end{itemize}

The rest of this paper is organized as follows: The system model is formulated in Section \ref{sm}. Section \ref{GM} presents the hierarchical Stackelberg game analysis. The personalized FRL solution is detailed in Section \ref{solution}, while Section \ref{PE} evaluates its performance. Finally, conclusions drawn in Section \ref{conclusion}.

\section{System Model}\label{sm}
In the considered SAGIN, the LEO satellite constellation serves multiple ground hotspots indexed by $n\in \mathcal{N}$, with $N=|\mathcal{N}|$, over a finite time period $T=|\mathcal{T}|$. Each hotspot is supported by a dedicated RIS-equipped UAV (UAV-RIS) $n\in \mathcal{N}$ relay hovering above it, establishing a one-to-one mapping between hotspots and UAV relays. Each RIS consists of $M=|\mathcal{M}|$ RIS elements indexed by $m\in\mathcal{M}$. At time $t\in\mathcal{T}$, the number of active users in hotspot $n$ is denoted by $K_{n}[t] = |\mathcal{K}_{n}[t]|$, where each user is indexed by $k \in \mathcal{K}_{n}[t]$. A subset of users is equipped with FAS, where each FAS-enabled device contains $ H \triangleq H_{1}\times H_{2} = |\mathcal{H}|$ ports arranged in a two-dimensional array, with $H_{1}$ and $H_{2}$	denoting the numbers of rows and columns, and each port indexed by $h\in\mathcal{H}$. 

In this section, we first present the spatial coordinate model of the considered SAGIN, including user distributions, UAV trajectories, and LEO satellite orbits. We then derive the end-to-end downlink channel model from the LEO satellite to ground users and specify the FRL framework. Finally, we formulate the long-term sum-rate maximization problem.

\subsection{Spatial Coordinate Model}\label{spatialM}

\begin{figure}[htbp]
\centering
\centerline{\includegraphics[scale=0.324]{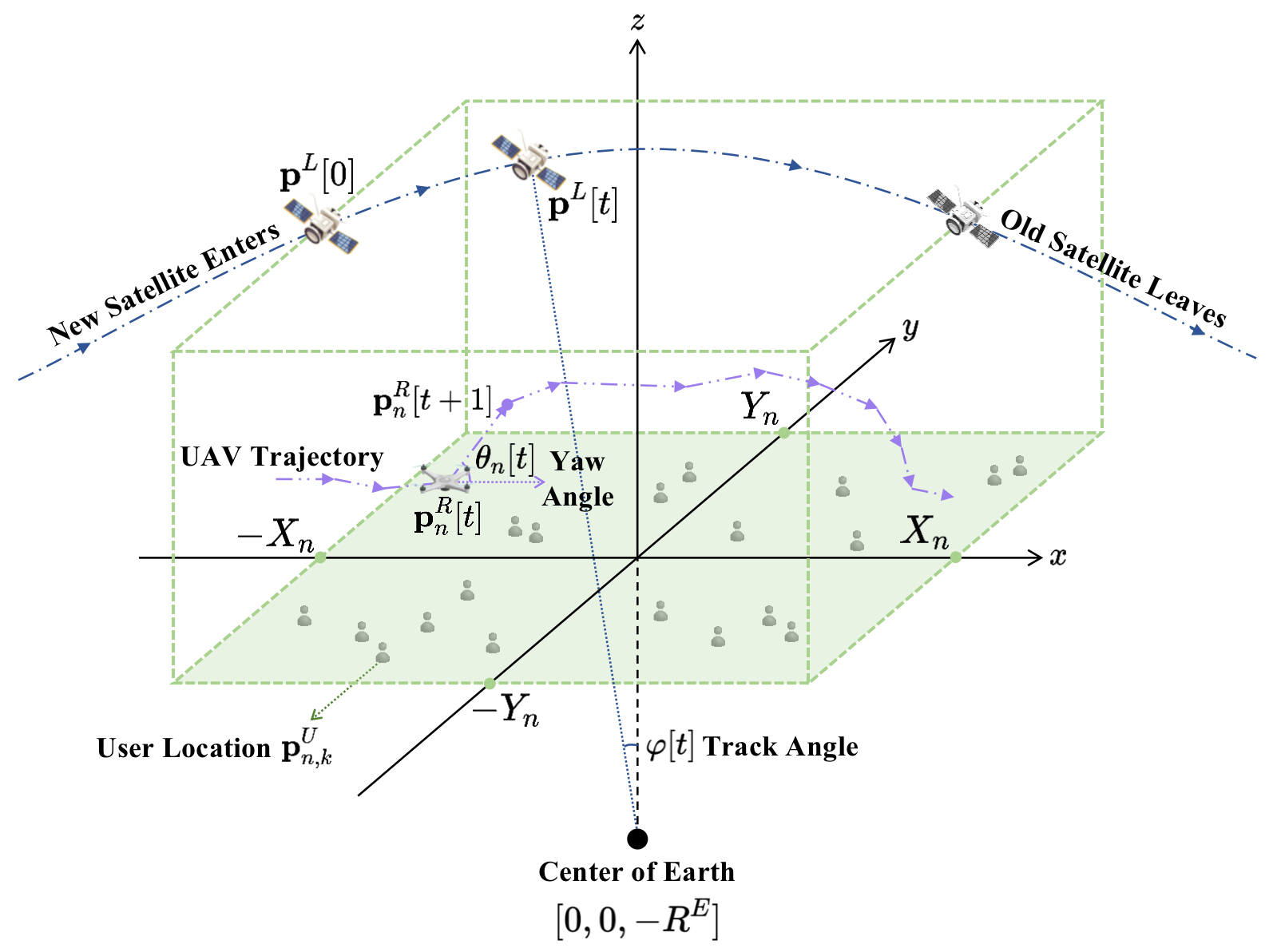}}
\caption{Illustration of spatial coordinates.}
\label{spatial}
\end{figure}

The system model involves three types of entities: ground users, UAV-RIS relays, and LEO satellite. 

\subsubsection*{\textbf{Users Distribution}}
For ground users, it is assumed that their positions remain fixed during the considered time period $T$. Accordingly, the spatial coordinate of user $(n,k)$ is represented by a constant vector $\mathbf{p}^{U}_{n,k}\triangleq[x_{n,k}^{U},y_{n,k}^{U},0]^\top\in\mathbb{R}^{3}$. As illustrated in Fig. \ref{spatial}, the coverage area of hotspot $n$ is defined as a rectangular region bounded by $[-X_{n}, X_{n}]$ along the $x$-axis and $[-Y_{n}, Y_{n}]$ along the $y$-axis\footnote{The rectangular region is assumed only for modeling convenience. The proposed problem and algorithm are not restricted to this particular geometry.}.

\subsubsection*{\textbf{UAVs Trajectory}}
For UAV-RIS relay $n$, its instantaneous spatial coordinate at time $t$ is defined as
$\mathbf{p}^{R}_{n}[t] \triangleq [x^{R}_{n}[t], y^{R}_{n}[t], z^{R}_{n}]^\top \in \mathbb{R}^{3}$,
where a fixed cruising altitude $z^{R}_{n}$ is assumed. The distance between relay $n$ and user $(n,k)$ at time $t$ is defined as
\begin{equation}\label{dist_RU}
    d_{n,k}^{RU}[t]=||\mathbf{p}^{R}_{n}[t]-\mathbf{p}^{U}_{n,k}||.
\end{equation}

In the horizontal $x–y$ plane, the UAV's flight direction is controlled by the yaw angle $\theta_{n}[t] \in [0, 2\pi)$, and the flight distance is denoted by $v_{n}[t]$. Accordingly, the relationship between the positions at two consecutive time is given by:
\begin{equation}\label{coord_relay}
    \mathbf{p}^{R}_{n}[t+1] = \mathbf{p}^{R}_{n}[t] +v_{n}[t]\cdot \mathbf{e}_{n}[t],
\end{equation}
where $\mathbf{e}_{n}[t]\triangleq [\cos(\theta_{n}[t]), \sin(\theta_{n}[t]),0]^\top$ is the unit direction vector. Due to the physical limitations of UAVs, the maximum allowable movement distance within a single time $t$ is constrained by $v_{n}^{\max}$, i.e., $v_{n}[t]\leq v_{n}^{\max}$, $\forall t \in T$.

\subsubsection*{\textbf{LEO Satellite Orbit}}
The spatial coordinate of the satellite at time $t$ is defined as $\mathbf{p}^{L}[t]\triangleq [x^{L}[t],y^{L},z^{L}[t]]^\top\in \mathbb{R}^{3}$. The vast majority of operational commercial and scientific LEO satellites, especially LEO constellations, are placed in prograde orbits, i.e., eastward, taking advantage of Earth's rotation to achieve orbital velocity more efficiently \cite{Capderou2005}. Similar to \cite{10702554} and \cite{10155303}, the satellite is assumed to move along the positive $x$-axis, and its altitude varies along the $z$-axis.

Specifically, the satellite follows a circular orbit in the $x–z$ plane, centered at the center of Earth located at $[0, 0, R^{E}]$, where $R^{E}$ denotes the Earth's radius. The angle between the satellite position vector (from the center of Earth) and the positive $z$-axis is referred to as the orbit angle, denoted by $\varphi[t] \in [0, 2\pi)$. The orbit angle evolves with angular velocity $\omega$, such that $\varphi[t] = \varphi[0] + \omega t$. Based on the above formulation, the satellite’s position at time $t$ is given by:
\begin{equation}\label{LEO_p}
    \mathbf{p}^{L}[t] = \begin{bmatrix}
x^{L}[t] = R^{L}\sin{(\varphi[0] + \omega t)} \\
y^{L} \\
z^{L}[t] = -R^{E}+R^{L}\cos{(\varphi[0] + \omega t)}
\end{bmatrix},
\end{equation}
and the distance from LEO satellite to UAV-RIS relay $n$ at time $t$ is defined by:
\begin{equation}
    d_{n}^{LR}[t]=||\mathbf{p}^{L}[t]-\mathbf{p}^{R}_{n}[t]||.
\end{equation}

Due to the high mobility of LEO satellites, a single satellite cannot serve continuously over the entire period $T$. Leveraging ultra-dense LEO constellations, such as SpaceX’s Starlink, temporal coverage continuity can be maintained through satellite handovers \cite{10228912}. Following the handover model in \cite{10702554}, multiple satellites may pass over hotspot $n$ during period $T$, while only one serves it at each time. As shown in Fig. \ref{spatial}, such continuity is ensured by seamlessly switching to a new satellite when the old one leaves the coverage area.

Accordingly, the orbit angles at which the satellite enters and leaves the area of hotspot $n$ are given by:
\begin{equation}
\varphi_{\text{enter}} = \arcsin(\frac{-X_{n}}{R^{L}} ), \
\varphi_{\text{leave}} = \arcsin(\frac{X_{n}}{R^{L}}).
\end{equation}
The corresponding angular span of the orbit and the associated service duration can be expressed as:
\begin{equation}
    \Delta\varphi=\varphi_{\text{leave}} - \varphi_{\text{enter}},\
    \Delta t= \frac{\Delta\varphi}{\omega}.
\end{equation}
Therefore, Equation $(\ref{LEO_p})$ can be reformulated as:
\begin{equation}
\mathbf{P}^{L}[t]=\begin{bmatrix}
R^{L}\sin(\varphi_{\text{enter}}+\omega\cdot (t\ \text{mod} \ \Delta t)) \\
y^{L} \\
-R^{E}+R^{L}\cos(\varphi_{\text{enter}}+\omega\cdot (t\ \text{mod} \ \Delta t))
\end{bmatrix},
\end{equation}
where $(t\ \text{mod} \ \Delta t)$ denotes the elapsed service time since the satellite entered the coverage area of the hotspot.

\subsection{Channel Model}\label{CM}

\begin{figure}[htbp]
\centering
\centerline{\includegraphics[scale=0.38]{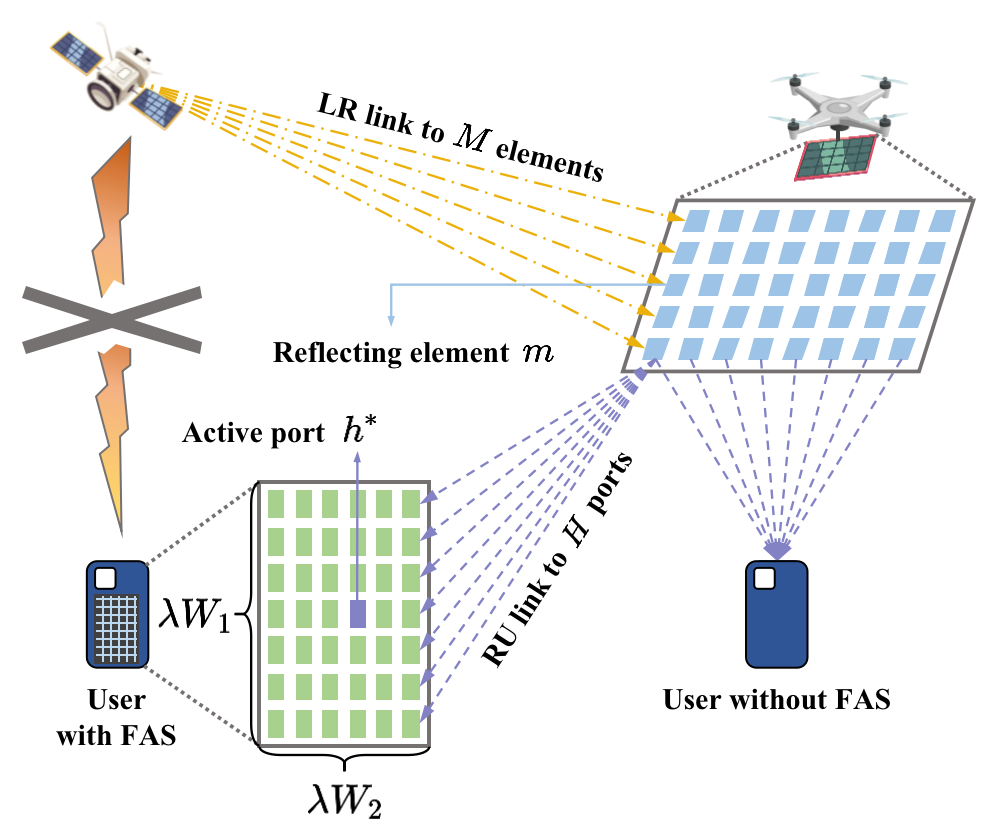}}
\caption{Illustration of communication channel.}
\label{channel}
\end{figure}

The direct link from LEO satellites to ground users is often affected by buildings or cloud coverage, resulting in masking and intermittent blockage\cite{9822386,10445520,10702554, 10207732}. Therefore, as illustrated in Fig. \ref{channel}, this work focuses on the reflected downlink, which is divided into two segments: the LEO-to-UAV-RIS link (LR) and the UAV-RIS-to-user link (RU).

\subsubsection*{\textbf{LEO-to-UAV-RIS Link}}

The channel between LEO satellite to UAV-RIS relay $n$ at time $t$ is defined by $\mathbf{h}^{LR}_{n}[t]\in \mathbb{C}^{M\times1}$, where each element represents the channel between the LEO satellite and the reflecting element $m \in \mathcal{M}$ of RIS $n$. A Rician fading model is adopted, which is commonly utilized in \cite{9822386,10702554,10848151}. The corresponding channel vector is defined as:
\begin{equation}
    \mathbf{h}^{LR}_{n}[t] = \frac{\lambda}{4\pi d^{LR}_{n}[t]\sqrt{K^{LR}+1}}(\sqrt{K^{LR}}\bar{\mathbf{h}}^{LR}+\mathbf{g}^{LR}_{n}[t]),
\end{equation}
where $K^{LR}$ is the Rician factor of LR link, and $\bar{\mathbf{h}}^{LR}$ is the LoS component. 

The LoS signal is typically strong and highly directional, particularly in far-field scenarios such as satellite–UAV links. Since the physical size of the RIS surface is small relative to the large propagation distance, the wavefront of the LoS signal can be approximated as parallel when it arrives at the RIS. As a result, the LoS component exhibits minimal variation across different RIS elements. Therefore, $\bar{\mathbf{h}}^{LR}$ is set to $\mathbf{1}_{M}$.
$\mathbf{g}^{LR}_{n}[t]\sim \mathcal{CN}(0,\mathbf{I}_{M})$ is the NLoS component. $\lambda = \frac{c}{f_{c}}$ is the wavelength, where $c$ is the light speed and $f_{c}$ is the frequency of carrier wave.

\subsubsection*{\textbf{UAV-RIS-to-user Link}}
Ground user equipment can be categorized into two types: those equipped with FAS (FAS users) and those without FAS capability. As illustrated in Fig.~\ref{channel}, if the user is equipped with FAS, each reflecting element $m$ of the RIS forms a RU link with each port $h$, resulting in a total of $M \times H$ such links. Therefore, the channel matrix $\mathbf{H}^{RU}_{n,k}[t]\in \mathbb{C}^{M\times H}$ of UAV-RIS relay $n$ to ground user $(n,k)$ at time $t$ is defined by:
\begin{equation}\label{RU_channel_matrix}
    \mathbf{H}^{RU}_{n,k}[t] = \frac{\lambda }{4\pi d^{RU}_{n,k}[t]\sqrt{K^{RU}+1}}(\sqrt{K^{RU}}\bar{\mathbf{H}}^{RU}+\mathbf{G}^{RU}_{n,k}[t]),
\end{equation}
where $K^{RU}$ is the Rician factor of RU link. $\bar{\mathbf{H}}^{RU}$ is the LoS component, commonly set to $\mathbf{1}_{M\times H}$.

$\mathbf{G}^{RU}_{n,k}[t]\in \mathbb{C}^{M\times H}$ is the NLoS component, which consists of signals reflected by the RIS that reach the user via multiple propagation paths. Due to the multipath effect’s sensitivity to spatial distribution, different FAS ports receive distinct superpositions of these reflected signals. The NLoS component exhibits significant variation across ports and shows noticeable spatial correlation, which is defined as follows:
\begin{equation}
    \text{vec}(\mathbf{G}^{RU}_{n,k}[t])\sim \mathcal{CN}(0,\mathbf{I}_{M}\otimes \mathbf{R}),
\end{equation} 
where $\text{vec}(\cdot)$ represents vector stacking, which pulls the $M \times H$ matrix into a column vector $MH \times 1$ to establish a joint distribution. $\otimes$ is the kronecker product. $\mathbf{R}\in \mathbb{R}^{H\times H}$ is the spatial correlation matrix among the ports, and expressed as:
\begin{equation}
    \mathbf{R} = \begin{bmatrix}
 \varpi_{1,1} & \cdots  & \varpi_{1,\tilde{h}} & \cdots & \varpi_{1,H}\\
  \vdots      &         & \vdots               &        & \vdots\\
 \varpi_{h,1} & \cdots  & \varpi_{h,\tilde{h}} & \cdots & \varpi_{h,H}\\
 \vdots       &         & \vdots               &        & \vdots\\
 \varpi_{H,1} & \cdots  & \varpi_{H,\tilde{h}} & \cdots & \varpi_{H,H}
\end{bmatrix}.
\end{equation}

Each element $\varpi_{h,\tilde{h}}$ is the spatial correlation of ports $h = \mathcal{F}^{-1}(h_{1},h_{2})$ and $\tilde{h} = \mathcal{F}^{-1}(\tilde{h}_{1},\tilde{h}_{2})$, which is defined by:
\begin{equation}
    \varpi_{h,\tilde{h}} = \text{sinc}(\frac{2}{\lambda}\sqrt{(\frac{|h_{1}-\tilde{h}_{1}|}{H_{1}-1}W_{1})^{2} + (\frac{|h_{2}-\tilde{h}_{2}|}{H_{2}-1}W_{2} )^{2}}),
\end{equation}
where $W_{1}$ and $W_{2}$ are the size parameters of FAS, i.e., $\lambda W_{1}$ and $\lambda W_{2}$ define the physical size the FAS planar\cite{10539238}. 

If the user is not equipped with FAS, the channel between UAV-RIS relay $n$ to ground user $(n,k)$ at time $t$, denoted by $\mathbf{h}^{RU\setminus F}_{n,k}[t]\in \mathbb{C}^{M\times 1}$, is defined by:
\begin{equation}
    \mathbf{h}^{RU\setminus F}_{n,k}[t] = \frac{\lambda }{4\pi d^{RU}_{n,k}[t]\sqrt{K^{RU}+1}}(\sqrt{K^{RU}}\bar{\mathbf{h}}^{RU}+\mathbf{g}^{RU}_{n,k}[t]),
\end{equation}
where $\bar{\mathbf{h}}^{RU}$ is the LoS component, which is set to $\mathbf{1}_{M}$.
$\mathbf{g}^{RU}_{n,k}[t]\sim \mathcal{CN}(0,\mathbf{I}_{M})$ is the NLoS component.

\subsection{Federated Settings}\label{FS}

As discussed in Section \ref{intro}, the FRL framework naturally fits the hierarchical architecture of SAGINs. In this work, each UAV-RIS relay acts as an agent that trains a local policy for joint UAV trajectory and RIS phase optimization, while the LEO satellite constellation functions not only as a communication entry but also as a global server that aggregates local models and updates a global policy.

Due to the high mobility of LEO satellites, the multi-satellite handover mechanism in Subsection \ref{spatialM} ensures service continuity. To support the FRL framework under such handovers, inter-satellite links (ISLs) are employed to enable model parameter aggregation and dissemination across the constellation \cite{9442378}. As ISLs facilitate efficient inter-satellite cooperation, they provide a practical means to maintain continuous federated training in multi-satellite SAGINs \cite{10646360}.

FRL relies on interaction–generated experience rather than static datasets. Therefore, each satellite can train the global model during its service period without data inheritance. When a satellite leaves the hotspot coverage, the global model parameters are transferred to the next serving satellite via ISLs and routing mechanisms, thereby ensuring training continuity of the FRL framework across the LEO constellation \cite{10409275}.

For simplification, the global server is abstracted as a fixed LEO satellite entity in the FRL framework. However, when modeling the actual communication channels, the realistic scenario of dynamic satellite handovers remains.

\subsection{Problem Formulation}\label{PF}
By combining the two channel segments modeled in Subsection \ref{CM}, the equivalent channel from the LEO satellite to user $(n,k)$ can be defined. 

If user $(n,k)$ is equipped with FAS, the active port $h^{*}$ that receives the optimal channel gain is defined as:
\begin{equation}
    h^{*}=\mathop{\arg\max}\limits_{h \in \mathcal{H}}|[\mathbf{h}^{LR}_{n}[t]^\top \Phi^{R}_{n}[t]\mathbf{H}^{RU}_{n,k}[t]]_{h}|^{2},
\end{equation}
where $[\cdot]_{h}$ denotes the $h$-th element. The RU channel corresponding to the active port $h^{*}$ is denoted as $\mathbf{h}^{RU}_{n,k,h^{*}}[t] \in \mathbb{C}^{M \times 1}$, representing the $h^{*}$-th column of $\mathbf{H}^{RU}_{n,k}[t]$. $\Phi^{R}_{n}[t]\in\mathbb{C}^{M\times M}$ is the phase shift matrix, which is defined as:
\begin{equation}
    \Phi^{R}_{n}[t] = \text{diag} ([e^{j\phi_{n,1}[t]},\cdots,e^{j\phi_{n,m}[t]},\cdots, e^{j\phi_{n,M}[t]}]),
\end{equation}
where $\phi_{n,m}[t]$ is the phase shift of element $m$ at time $t$.

Based on the above discussion, the equivalent channel can be expressed in a unified form as:
\begin{equation}\label{h_RU}
    h_{n,k}[t]=\left\{\begin{matrix}
\mathbf{h}^{LR}_{n}[t]^\top \Phi^{R}_{n}[t]\mathbf{h}^{RU}_{n,k,h^{*}}[t],       & \text{if with FAS}, \\
\mathbf{h}^{LR}_{n}[t]^\top \Phi^{R}_{n}[t]\mathbf{h}^{RU\setminus F}_{n,k}[t],  & \text{if without FAS}.
\end{matrix}\right.
\end{equation}

In the considered communication scenario, the satellite broadcasts to all users within the same hotspot. Alternatively, as discussed in Subsection \ref{FS}, the satellite constellation can also broadcast different signals across hotspots. Accordingly, the downlink transmission rate of user $(n,k)$ is given by:
\begin{equation}\label{rate}
    r_{n,k}[t]=B\log_{2}(1+\bar{\gamma}|h_{n,k}[t]|^{2}),
\end{equation}
where $B$ is the channel bandwidth, and $\bar{\gamma}= \frac{\bar{P}}{\sigma^{2}}$ is the average SNR. Here, $\bar{P}$ denotes the transmit power with equivalent isotropically radiated power (EIRP) spectral density $\text{PSD}_{\text{EIRP}}$ [dBW/4kHz], which includes the transmit antenna gain \cite{fcc2020spacex}. $\sigma^{2}$ is the power of the additive white Gaussian noise (AWGN) with zero mean and power spectral density $N_{0}$ [dBm/Hz].

Accordingly, the long-term sum-rate maximization problem for all users can be formulated as:

\begin{align}
\max_{\mathbf{p}^{R}_{n}[t],\Phi^{R}_{n}[t]} \quad & \sum_{t\in \mathcal{T}} \sum_{n\in \mathcal{N}} \sum_{k\in \mathcal{K}_{n}[t]} r_{n,k}[t], \quad \forall n \in N \label{OP_obj} \\
\text{s.t.} \quad 
& \text{C1: } -X_{n} \leq x^{R}_{n}[t] \leq X_{n}, \tag{\ref{OP_obj}{a}}\label{OP:c1} \\
& \text{C2: } -Y_{n} \leq y^{R}_{n}[t] \leq Y_{n}, \tag{\ref{OP_obj}{b}}\label{OP:c2} \\
& \text{C3: } 0 \leq v_{n}[t] \leq v_{n}^{\max}, \tag{\ref{OP_obj}{c}}\label{OP:c3} \\
& \text{C4: } 0 \leq \theta_{n}[t] < 2\pi, \tag{\ref{OP_obj}{d}}\label{OP:c4} \\
& \text{C5: } \phi_{n,m}[t] = \frac{2\pi(c-1)}{C}. \tag{\ref{OP_obj}{e}}\label{OP:c5}
\end{align}

Constraints \eqref{OP:c1} and \eqref{OP:c2} define the horizontal flight boundaries of UAV $n$ within the coverage area. Constraints \eqref{OP:c3} and \eqref{OP:c4} impose restrictions on the flight actions of UAV $n$ at each time $t$.
Constraint \eqref{OP:c5} specifies a discrete phase control, where the feasible phase values are uniformly quantized into $C$ discrete points, and $c \in C$.

Equation \eqref{h_RU} implicitly involves binary variables indicating whether a user is equipped with FAS and activated states, rendering the optimization problem in \eqref{OP_obj} a mixed-integer nonlinear programming (MINLP) problem, which is challenging to solve due to its combinatorial and nonconvex nature. Moreover, as modeled in Subsection \ref{CM}, both channel segments incorporate time-varying random components. Consequently, solving the optimization problem in \eqref{OP_obj} requires capturing the long-term statistical characteristics of the channel to ensure efficient decision-making.

\section{Hierarchical Stackelberg Game}\label{GM}
LEO satellites, UAV relays, and ground users in SAGINs inherently exhibit leader–follower characteristics interaction. In particular, UAV decisions influence user behavior, while the LEO satellite guides UAV policy learning under the FRL framework. Motivated by this hierarchical structure, the optimization problem in \eqref{OP_obj} is decomposed into two levels and modeled as a Stackelberg game, respectively, which enables tractable analysis and facilitates effective algorithm design.

\subsection{Stackelberg Game of UAV Relay and Users}
Let $\hat{\mathcal{K}}_{n}[t] \subseteq \mathcal{K}_{n}[t]$ denote the set of users equipped with FAS within hotspot $n$ at time $t$. A Stackelberg game is formulated between UAV-RIS relay $n \in N$ and its associated FAS-equipped users $k \in \hat{\mathcal{K}}_{n}[t]$, where the UAV-RIS relay acts as the leader and the users as followers. This structure is demonstrated by the fact that the UAV’s position $\mathbf{p}^{R}_{n}[t]$ and the RIS phase control matrix $\Phi^{R}_{n}[t]$ mainly determine the channel, based on which each user determines the active port.

\begin{definition}[Stackelberg Game $\mathcal{G}^{RU}_{n,[t]}$]\label{defi_game_RU}
    At each time slot $t$, UAV-RIS relay $n$ influences the port activation decisions of FAS-equipped users $k \in \hat{\mathcal{K}}_{n}[t]$ through its trajectory and RIS phase control. This interaction is modeled as a Stackelberg game $\mathcal{G}^{RU}_{n,[t]}$, defined as follows:
    \begin{equation}
        \mathcal{G}^{RU}_{n,[t]}=\left \langle \{n\}\cup \hat{\mathcal{K}}_{n}[t],\{\mathcal{A}^{R}_{n},\mathcal{A}^{U}_{n,k}\},\{U^{R}_{n}[t],u^{U}_{n,k}[t]\}  \right \rangle,  
    \end{equation}
    where ${\mathcal{A}^{R}_{n}, \mathcal{A}^{U}_{n,k}}$ denote the action spaces of UAV-RIS relay $n$ and user $(n,k)$, respectively, while ${U^{R}_{n}[t], u^{U}_{n,k}[t]}$ represent their corresponding utilities at time $t$.
\end{definition}

The action of UAV-RIS relay $n$ consists of its trajectory and phase shift control, and is thus defined as 
\begin{equation}\label{relay_action}
    a^{R}_{n}[t] = \{\theta_{n}[t],v_{n}[t],\{\phi_{n,m}[t]\}_{m\in M}\}\in \mathcal{A}^{R}_{n}.
\end{equation}
The action of user $(n,k)$ is reactive, meaning it is determined in response to the UAV-RIS relay’s action. It is defined as
\begin{equation}\label{user_action}
    a_{n,k}^{U}[t] = \mathop{\arg\max}\limits_{h \in \mathcal{H}}|f_{h}(a_{n}^{R}[t])|^{2}\in \mathcal{A}^{U}_{n,k},
\end{equation}
where $f_{h}(a_{n}^{R}[t])$ denotes the equivalent channel determined by action $a^{R}_{n}[t]$ at port $h$, which is defined as
\begin{equation}\label{eq_channel_over_a}
    f_{h}(a_{n}^{R}[t])=\mathbf{h}^{LR}_{n}(a)^\top \Phi_{n}^{R}(a)\mathbf{h}^{RU}_{n,k,h}(a)|_{a=a_{n}^{R}[t]}.
\end{equation}

The utility of user $(n,k)$ at time $t$ is  defined by its rate, i.e.,
\begin{equation}
u_{n,k}^{U}[t] = r_{n,k}[t], \quad \forall k \in \hat{\mathcal{K}}_{n}[t].
\end{equation}
The utility of UAV-RIS relay $n$ is defined as the sum-rate of all users in the associated hotspot:
\begin{equation}\label{utility_R}
U_{n}^{R}[t] = \sum_{k \in \mathcal{K}_{n}[t]} r_{n,k}[t].
\end{equation}
It is worth noting that this utility includes the rates of users without FAS. These users passively respond to the relay’s action $a_{n}^{R}[t]$, and their rates are thus considered part of the relay’s overall utility. Accordingly, the NE of this game can be defined as follows:
\begin{definition}[NE of game $\mathcal{G}^{RU}_{n,[t]}$]\label{defi_NE_game_RU}
    A strategy profile $(a^{R*}_{n}[t], \{ a^{U*}_{n,k}[t] \}_{k\in\hat{\mathcal{K}}_{n}[t]})$ constitutes a NE of the game $\mathcal{G}^{RU}_{n,[t]}$ if no player—neither the relay $n$ nor any user $k \in \hat{\mathcal{K}}_{n}[t]$—has an incentive to unilaterally deviate from their strategy given the strategies of the others.
\end{definition}
The following theorem is based on Definitions \ref{defi_game_RU} and \ref{defi_NE_game_RU}.
\begin{theorem}\label{theo_NE_game_RU}
    At each time $t$, the Stackelberg game $\mathcal{G}^{RU}_{n,[t]}$ among UAV-RIS relay $n$ and users $k \in \hat{\mathcal{K}}_{n}[t]$ admits at least one NE.
\end{theorem}
\begin{prooff}
    Please refer to Appendix A.
\end{prooff}

\subsection{Stackelberg Game of LEO Satellite and UAVs}\label{SG_LU}

As illustrated in Fig. \ref{ssm}, the LEO satellite acts as a global server (global node) and forms a FRL framework with multiple UAV-RIS relays (local nodes). At each time $t$, the LEO satellite aggregates local policy parameters to generate global policy parameters $\vartheta^{L}_{[t]}\in\Theta^{L}$ and broadcasts it to all relays. Each UAV relay then applies a local mapping to obtain local policy parameters $\vartheta^{R}_{n,[t]}\in\Theta^{R}_{n}$. The mutual mapping between gloabl and local parameters can be defined as follows:
\begin{equation}
    \vartheta^{R}_{n,[t]} = \mathcal{F}^{L\to R}(\vartheta^{L}_{[t]}),
\end{equation}
\begin{equation}
    \vartheta^{L}_{[t]} = \mathcal{F}^{R\to L}(\vartheta^{R}_{n,[t]}).
\end{equation}

Based on these parameters, both the global and local nodes can construct corresponding policies and subsequently generate actions. Specifically, we have
\begin{equation}
    a_{n}^{R}[t]\sim\pi_{\vartheta_{n,[t]}^{R}}(\cdot),\ a^{L}[t]\sim\pi_{\vartheta_{[t]}^{L}}(\cdot),
\end{equation}
where $(\cdot)$ represents the input of state. Under the assumption of complete information, both the global policy $\pi_{\vartheta^{L}_{[t]}}$ and the local policies $\pi_{\vartheta^{R}_{n,[t]}}$ share complete state inputs. Therefore, the following Stackelberg game is defined.

\begin{definition}[Stackelberg Game $\mathcal{G}^{LR}_{[t]}$]\label{defi_game_LR}
    At each time slot $t$, LEO satellite acts as a global server influences the action generation policy of all UAV-RIS relays through global parameters. This interaction is modeled as a Stackelberg game $\mathcal{G}^{LR}_{[t]}$, defined as follows:
    \begin{equation}
        \mathcal{G}^{LR}_{n,[t]}=\left \langle \{l\}\cup \mathcal{N},\{\Theta^{L},\Theta^{R}_{n}\},\{U^{L}[t],U^{R}_{n}[t]\}  \right \rangle,  
    \end{equation}
    where $l$ denotes the LEO satellite serving hotspot $n$. $\{\Theta^{L},\Theta^{R}_{n}\}$ denote the action spaces of LEO satellite and UAV-RIS relay $n$, respectively, while $\{U^{L}[t],U^{R}_{n}[t]\}$ represent their corresponding utilities at time $t$.
\end{definition}

The utility of each UAV-RIS relay follows Eq. \eqref{utility_R}, while the utility of the LEO satellite is defined as the sum of all relay utilities:
\begin{equation}
    U^{L}[t]=\sum_{n\in \mathcal{N}}U^{R}_{n}[t].
\end{equation}
Note that each $U^{R}_{n}[t]$ is computed based on the virtual action $a^{L}[t]$ generated by the satellite, which is applied across all relays to evaluate the resulting utility. 

The NE of the Stackelberg game of LEO satellite and UAVs can be defined as follows:
\begin{definition}[NE of game $\mathcal{G}^{LR}_{[t]}$]\label{defi_NE_game_LR}
    A strategy profile $(\vartheta^{L*}_{[t]}, \{ \vartheta^{R*}_{n,[t]} \}_{n\in \mathcal{N}}$ constitutes a NE of the game $\mathcal{G}^{LR}_{[t]}$ if no player—neither the LEO satellite $l$ nor any relay $n \in \mathcal{N}$—has an incentive to unilaterally deviate from their strategy given the strategies of the others.
\end{definition}
The following theorem is based on Definitions \ref{defi_game_LR} and \ref{defi_NE_game_LR}.
\begin{theorem}\label{theo_NE_game_LR}
    At each time $t$, the Stackelberg game $\mathcal{G}^{LR}_{[t]}$ among LEO satellite $l$ and relays $n\in \mathcal{N}$ admits at least one NE.
\end{theorem}
\begin{prooff}
    Please refer to Appendix B.
\end{prooff}

\subsection{Markov Game Formulation}\label{MGF}
Through the hierarchical Stackelberg game analysis, the proposed optimization problem in \eqref{OP_obj} is demonstrated to be solvable. To facilitate the design of an effective solution, the hierarchical Stackelberg game can be reformulated as a Markov game, where each UAV-RIS relay is treated as an agent. The Markov game is defined by a tuple $\left\langle \mathcal{N},\mathcal{S},\mathcal{A},\mathbf{P},\mathbb{R},\eta \right\rangle$. Specifically, $N$ is the set of agents.
    
State space is defined by $\mathcal{S} = \{\mathbf{s}_{n}[t]\}_{n\in \mathcal{N}, t\in \mathcal{T}}$, where $\mathbf{s}_{n}[t]$ is the state of agent $n$ at time $t$. A fixed state dimension is ensured by introducing the activation indicator $\nu_{n,k}[t]$ to model user activity. Let $\mathcal{K}_{n}$ denote the set of all users in hotspot $n$, with $K_{n}= |\mathcal{K}_{n}|$. The number of active users at time $t$ is then given by $K_{n}[t]=\sum_{k\in\mathcal{K}_{n}}\nu_{n,k}[t]$. In addition, $q_{n,k}[t]$ indicates whether a user is equipped with FAS. Therefore, $$\mathbf{s}_{n}[t]\triangleq \left \langle \mathbf{p}^{L}[t],\mathbf{p}^{R}_{n}[t-1],\{\mathbf{p}^{U}_{n,k},\nu_{n,k}[t],q_{n,k}[t]\}_{k\in \mathcal{K}_{n}} \right \rangle,$$ including the coordinates of LEO satellite, UAV and all ground users, along with the activation and FAS states of each user.

Action space is defined as the union of the action spaces of all agents $\mathcal{A}\triangleq \bigcup_{n \in \mathcal{N}}\mathcal{A}^{R}_{n} $, where the action of each agent at time $t$ is defined in Eq. \eqref{relay_action}.

$\mathbf{P}: \mathcal{S}\times\mathcal{A}\rightarrow \Delta(\mathcal{S})$ is the state transition probability distribution, where $\mathbf{P}(s'|s,a)\triangleq \Pr(\mathbf{s}_{n}[t+1] \mid \mathbf{s}_{n}[t], a^{R}_{n}[t])$ represents the probability that agent $n$ transitions from state $\mathbf{s}_{n}[t]$ to $\mathbf{s}_{n}[t+1]$ after taking action $a^{R}_{n}[t]$.

The reward of each agent $n$ at time $t$ is defined by $r_{n}[t] = \frac{1}{K_{n}[t]}U_{n}^{R}[t]: \mathcal{S} \times \mathcal{A} \to \mathbb{R}^{+}$, i.e., the average transmission rate of active users in hotspot $n$, which mitigates deviations introduced by fluctuations in the number of active users over time. The total reward is obtained by accumulating average rates over the period $T$, i.e., $R_{n} = \sum_{t\in \mathcal{T}}r_{n}[t]$.

\section{Personalized Federated Reinforcement Learning Solution}\label{solution}

\begin{figure*}[t]
		\centering
		\includegraphics[scale=0.45]{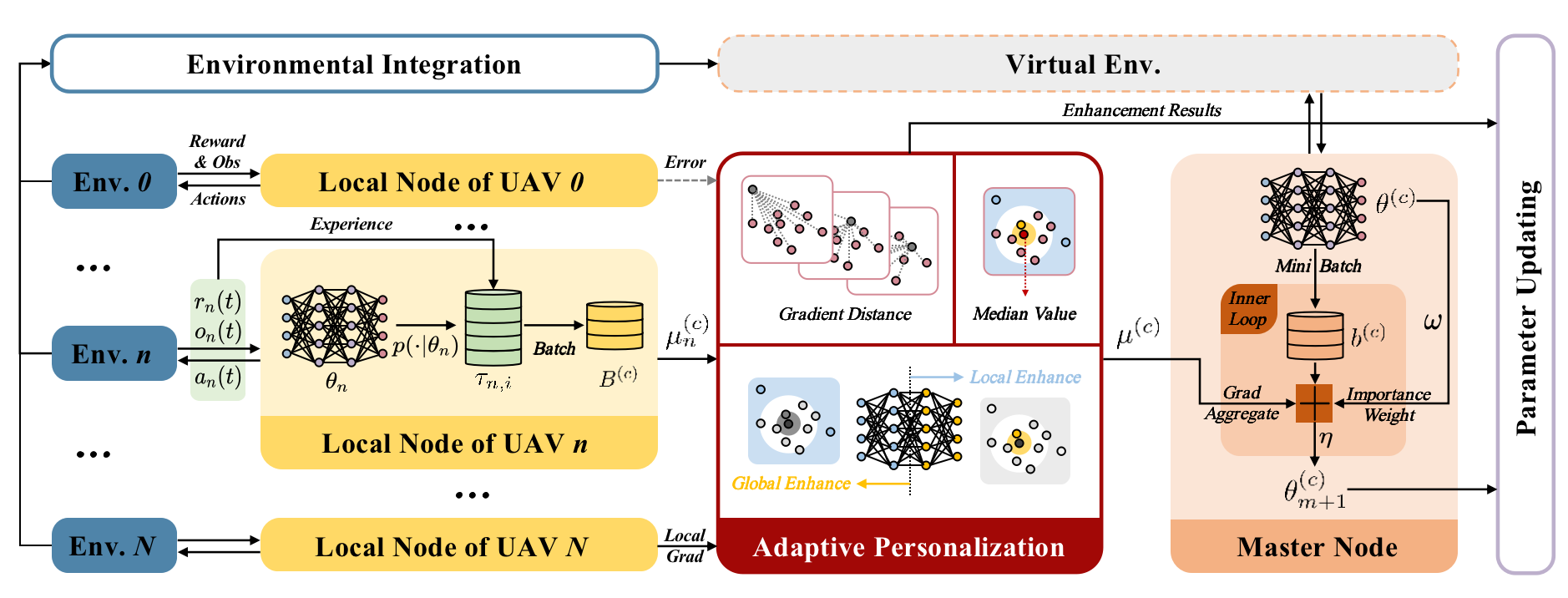}
		\renewcommand{\figurename}{Fig.}
		\caption{The training workflow of FedPG-AP.}
		\label{FedPG-AP}
	\end{figure*}

\begin{algorithm}[ht]
\caption{Training of FedPG-AP}
\label{alg:FedPG-AP}

\KwInit{

Policy parameters:
$\{\vartheta^{R}_{n,(0)}\}_{n\in\mathcal{N}}$, $\vartheta^{L}_{(0)}$;

Training hyperparameters:
batch sizes $B^{(0)}, b^{(0)}$;
initial partition $e_{0}$;
thresholds $\sigma_{\text{close}}, \sigma_{\text{far}}$;
step size $\eta$
}

\For{\textup{epoch} $c = 1$ \KwTo $C$}{
    $\widetilde{\vartheta}^{L} \gets \vartheta^{L}_{(c-1)}$

    \For{\textup{local node} $n \in \mathcal{N}$}{
        \For{\textup{layer} $z = 0$ \KwTo $Z-1$}{
            \eIf{$z < e_{n}^{(c-1)}$}{
                $\vartheta_{n,(c)}^{R}[z] \gets \vartheta_{n,(c-1)}^{R}[z]$
            }{
                $\vartheta_{n,(c)}^{R}[z] \gets \vartheta_{(c-1)}^{L}[z]$
            }
        }

        \For{$i = 1$ \KwTo $B^{(c)}$}{
            Sample experience $\tau_{n,i} \sim p(\cdot \mid \vartheta_{n,(c)}^{R})$
        }

        $\mu_{n}^{(c)} \gets \widehat{\nabla} J(\vartheta^{R}_{n,(c)})$
        \tcp*[r]{see Eq.~\eqref{localG}}
    }

    Get $\{\Delta e_{n}^{(c)}\}_{n\in \mathcal{N}}$ according to Algorithm~\ref{alg:ap}

    $\mu^{(c)} \gets \frac{1}{N}\sum_{n\in \mathcal{N}}\mu_{n}^{(c)}$

    Sample $M^{(c)} \sim \mathrm{Geom}\!\left(\frac{B^{(c)}}{B^{(c)}+b^{(c)}}\right)$

    \For{\textup{inner loop} $m = 0$ \KwTo $M^{(c)}-1$}{
        \For{$j = 1$ \KwTo $b^{(c)}$}{
            Sample experience $\tau_{j} \sim p(\cdot \mid \vartheta^{L}_{m,(c)})$
        }

        $\kappa_{m}^{(c)} \gets
        \widehat{\nabla} J(\vartheta^{L}_{m,(c)})
        -
        \widehat{\nabla} J(\widetilde{\vartheta}^{L})$
        \tcp*[r]{see Eq.~\eqref{IW}}

        $v_{m}^{(c)} \gets \kappa_{m}^{(c)} + \mu^{(c)}$

        $\vartheta^{L}_{m+1,(c)} \gets \vartheta^{L}_{m,(c)} + \eta v_{m}^{(c)}$
    }

    $\vartheta^{L}_{(c)} \gets \vartheta^{L}_{M^{(c)},(c)}$
}
\end{algorithm}

To address the Markov game in a dynamic cross-hotspot environment, an FRL framework enables experience sharing across hotspots. However, the environmental state is not only time-varying but also heterogeneous across hotspots, due to differences in user distributions, FAS user ratios, and activation probabilities. Consequently, conventional federated mechanisms may fail to capture hotspot-specific characteristics.

To address this challenge, a novel personalized FRL algorithm, termed as Federated Policy Gradient with Adaptive Personalization (FedPG-AP), is proposed. As shown in Fig. \ref{FedPG-AP}, the proposed method consists of three main modules: local training and gradient estimation, adaptive personalization (AP), and global training with gradient aggregation. Algorithm \ref{alg:FedPG-AP} presents the complete training procedure of FedPG-AP.

\subsection{Local Training}

As defined in Subsection \ref{SG_LU}, the policy parameters of each UAV relay (agent) are denoted by $\vartheta^{R}_{n} \in \Theta^{R}_{n}$. As shown in Fig. 4, each agent $n$ corresponds to a local node that interacts with local environment to collect experience and updates parameters via gradient descent. Local training is performed in an epoch-wise manner. In each epoch $c$, a batch of experience with size $B^{(c)}$ is collected\footnote{The epoch index (c) is distinguished from the system time index [t] by different brackets.}. Specifically,

\begin{equation}
    \tau_{n,i}=\{\mathbf{s}_{n}[t], a^{R}_{n}[t],r_{n}[t], \mathbf{s}_{n}[t+1]\}_{t\in \mathcal{T}}
\end{equation} 
denotes an experience trace over time period $T$, where $i\in \{1,\dots,B^{(c)}\}$. Once collecting $B^{(c)}$ traces, local node $n$ estimates the gradient based on the current parameters $\vartheta^{R}_{n,(c)}$,
\begin{equation}\label{localG}
    \mu_{n}^{(c)}=\frac{1}{B^{(c)}}\sum_{i=1}^{ B^{(c)}}g(\tau_{n,i}|\vartheta^{R}_{n,(c)}),
\end{equation}
where $g(\tau_{n,i}|\vartheta^{R}_{n,(c)})$ denotes the gradient computed from experience $\tau_{n,i}$ under the policy parameters of the current epoch.

To mitigate gradient estimation bias, the GPOMDP estimator \cite{baxter2001infinite} is employed, which is given by:
\begin{equation}
    g(\tau_{n,i}|\vartheta^{R}_{n,(c)}) = - \nabla\sum_{t\in \mathcal{T}} A_{n}[t] \cdot \log \pi_{\vartheta^{R}_{n,(c)}}(a^{R}_{n}[t]|\mathbf{s}_{n}[t]).
\end{equation}
Here, $A[t]$ denotes the advantage of executing action $a^{R}_{n}[t]$, while the logarithmic term represents the log-probability of sampling action $a^{R}_{n}[t]$ from the policy $\pi_{\vartheta^{R}_{n,(c)}}$ given state $\mathbf{s}_{n}[t]$. The advantage measures the accumulated future reward following an action, defined as:

\begin{equation}
    \hat{R}_{n}[t] = \sum_{t'=t}^{T}\gamma^{t'-t}r_{n}[t'],
\end{equation}
where $\gamma\in(0,1)$ is the discount factor that controls the decay of future rewards over time. To further reduce estimation error, the advantage is defined in a normalized form:
\begin{equation}
    A_n[t] = \!\left [\hat{R}_n[t] - \mathrm{mean}_{t \in \mathcal{T}}\left(\hat{R}_n[t]\right)\right ]/
{\mathrm{std}_{t \in \mathcal{T}}\left(\hat{R}_n[t]\right)}.
\end{equation}

\subsection{Adaptive Personalization}\label{AP}

\begin{algorithm}[t]
\caption{Adaptive Personalization}
\label{alg:ap}

\KwIn{Local gradients $\{\mu_{n}^{(c)}\}_{n\in N}$;
      Thresholds $\sigma_{\text{far}}, \sigma_{\text{close}}$}
\KwOut{Enhancement adjustment $\{\Delta e_{n}^{(c)}\}_{n\in \mathcal{N}}$}

\For{$n \in \mathcal{N}$}{
    \For{$n' \in \mathcal{N}$}{
        $d_{n,n'}^{(c)} \gets \lVert \mu_{n}^{(c)} - \mu_{n'}^{(c)} \rVert_2$\;
    }
}

\For{$n \in \mathcal{N}$}{
    $\displaystyle m_{n}^{(c)} \gets \Median_{n' \in \mathcal{N}\setminus n}\, d_{n,n'}^{(c)}$\;
}

$\displaystyle \bar{n} \gets \mathop{\arg\min}\limits_{n\in \mathcal{N}} m_{n}^{(c)}$\;

\For{$n \in \mathcal{N}$}{
    \If{$d_{n,\bar{n}}^{(c)} > \sigma_{\text{far}}$}{
        $\Delta e_n^{(c)} \gets -1$\tcp*{Global Enhance}
    }
    \If{$d_{n,\bar{n}}^{(c)} < \sigma_{\text{close}}$}{
            $\Delta e_n^{(c)} \gets 1$\tcp*{Local Enhance}
        }
    $\Delta e_n^{(c)}$\;
}

\Return{$\{\Delta e_1^{(c)}, \dots, \Delta e_N^{(c)}\}$}

\end{algorithm}

Personalization in FRL can be achieved through various mechanisms. Inspired by \cite{10.1145/3605573.3605641}, we adopt a network partitioning strategy in which different layers inherit either local or global parameters. Unlike \cite{10.1145/3605573.3605641}, we assign input-side layers as local layers to capture hotspot-specific characteristics, while output-side layers share global parameters to exploit common experience across agents. The partition layer is indexed by $e_{0}$. Moreover, stochastic and time-varying hotspot heterogeneity may render fixed partitioning ineffective, which motivates the proposed AP mechanism.

Algorithm \ref{alg:ap} details the AP in Fig. \ref{FedPG-AP}. Local gradients are first gathered, and their distances $d_{n,n'}^{(c)},\forall n,n'\in N$ are computed to measure policy divergence across agents. The median node $\bar{n}$ is then used as a reference to quantify this divergence. To identify the median node, the median distance between each node $n \in \mathcal{N}$ and all other nodes $n' \in \mathcal{N}\setminus{n}$ is first computed, denoted by $m_n^{(c)}$. The node with the smallest median distance $m_n^{(c)}$ is then selected as the median node $\bar{n}$.

The gradient distance $d^{(c)}_{n,\bar{n}}$ between node $n$ to the median node $\bar{n}$ determines whether enhancement is applied. Two thresholds $\sigma_{\text{close}}<\sigma_{\text{far}}$ are introduced. If $d^{(c)}_{n,\bar{n}}<\sigma_{\text{close}}$, local enhancement is applied by adding one local layer ($\Delta e_{n}^{(c)}\gets 1$). If $d^{(c)}_{n,\bar{n}}>\sigma_{\text{far}}$, global enhancement is applied by adding one global layer ($\Delta e_{n}^{(c)}\gets -1$). Otherwise, no adjustment is made.

The enhancement results at epoch $c$ is defined as an adjustment to the initial partition, i.e., $e_n^{(c)} = e_0 + \Delta e_n^{(c)}$, which is applied to parameter inheritance in the subsequent epoch. The adjustment $\Delta e_n^{(c)}$ is reset at the next epoch, returning the partition to $e_0$. This design enables targeted adaptation within each epoch while eliminating influence across epochs, thereby preserving adaptability to time-varying heterogeneous environments.

\subsection{Global Training}

A master node is used to train the global policy with the same architecture as the local one. To address hotspot heterogeneity as shown in Fig. \ref{FedPG-AP}, a virtual environment is constructed by integrating local environments. Statistical features of local states are extracted, including user coordinate distributions, FAS-user ratios, and activation probabilities, from which virtual environmental distributions are constructed with perturbations, and concrete states are sampled for interaction.

Similar to the local nodes, the master node interacts with the virtual environment and collects experience traces $\tau_{j},j\in \{1,\dots,b^{(c)}\}$, with batch size $b^{(c)} \ll B^{(c)}$ for global gradient estimation. To reduce estimation bias, Stochastic Variance-Reduced Policy Gradient (SVRPG) is adopted \cite{papini2018stochastic}. This method introduces an inner loop to iteratively refine the global gradient estimate, where each iteration is indexed by $m\in \{0,\dots,M^{(c)}-1\}$. The gradient corresponding to each inner iteration $m$ is defined as
\begin{equation}\label{globalG}
    v_{m}^{(c)} = \mu^{(c)} + \kappa_{m}^{(c)},
\end{equation}
where $\mu^{(c)} \triangleq\frac{1}{N}\sum_{n\in \mathcal{N}}\mu_{n}^{(c)}$ is the aggregation of local gradients, and $\kappa_{m}^{(c)}$ is the importance-weighted gradient.

Specifically, importance weighting reuses past gradient estimates by scaling them with the likelihood ratio between the current and previous policy distributions, based on which $\kappa_{m}^{(c)}$ is defined as:

\begin{equation}\label{IW}
    \kappa_{m}^{(c)} = \frac{1}{b^{(c)}}\sum_{j=1}^{ b^{(c)}}[g(\tau_{j}|\vartheta^{L}_{m,(c)})-w(\tau_{j}|\vartheta^{L}_{m,(c)},\widetilde{\vartheta}^{L})g(\tau_{j}|\widetilde{\vartheta}^{L})].
\end{equation}
Here, $g(\tau_{j}|\vartheta^{L}_{m,(c)})$ denotes the gradient computed from experience $\tau_{j}$ under the policy parameters $\vartheta^{L}_{m,(c)}$ at inner iteration $m$ of epoch $c$. $g(\tau_{j}|\widetilde{\vartheta}^{L})$ denotes the gradient evaluated under the policy parameters from the previous epoch, i.e., $\widetilde{\vartheta}^{L}\gets \vartheta^{L}_{(c-1)}$. $w(\tau_{j}|\vartheta^{L}_{m,(c)},\widetilde{\vartheta}^{L}) \triangleq p(\tau_{j}|\widetilde{\vartheta}^{L})/p(\tau_{j}|\vartheta^{L}_{m,(c)})$ is an importance weight from policy $\pi_{\vartheta^{L}_{m,(c)}}$ to previous policy $\pi_{\widetilde{\vartheta}^{L}}$. The master node iteratively updates the policy parameters based on $v^{(c)}_{m}$ until the inner loop terminates.

Finally, the global parameters $\vartheta^{L}_{(c)}$ are broadcast to each local node, which updates its policy according to the enhancement result $e_{n}^{(c)}$. Let $z \in\{0, Z-1\}$ denote the network layer index. The $z$-th layer inherits local parameters for $z < e_n^{(c)}$ and global parameters otherwise. During execution, the computational complexity of FedPG-AP is $\mathcal{O}(T(\sum_{z=1}^{Z-1}n_{z}\cdot n_{z-1}))$, where $n_{z}$ denotes the number of neurons in the $z$-th layer. 

\section{Performance Evaluation}\label{PE}

\subsection{Settings}\label{Settings} 

The simulation follows the settings as below. A total of $N=5$ UAVs serve their corresponding hotspots over $T=30$ time units. Each hotspot $n$ lies within a square area bounded by $[-500, 500]$m in both dimensions $X_{n}$ and $Y_{n}$.

To introduce heterogeneity across hotspots, the user distribution, the proportion of FAS users, and the activation probability are perturbed around a common baseline. Each hotspot contains 10 randomly placed users, with a baseline FAS-user ratio of 0.5 and an activation probability of 0.8. Heterogeneity is achieved by assigning different center locations and clustering levels for user coordinates, along with random variations in FAS ratios and activation probabilities.

The FAS user device is equipped with $H = 25$ ports arranged in a $5 \times 5$ grid, with physical dimensions $W_1=W_2=3$ \cite{10539238}. The initial taking off position of each UAV is $[0,0,100]$, i.e., maintaining 100m cruising altitude \cite{10771971}. Each UAV carries an RIS comprising $M = 120$ reflecting elements, each supporting $C=50$ discrete phase levels.  

The real satellite constellation of the SpaceX Non-Geostationary Satellite System is adopted \cite{fcc2020spacex}. According to the 2020 FCC International Bureau Filings report, SpaceX applied to relocate 1,584 satellites to an altitude of 550 km. The Ku-band spectrum is allocated for the satellite-to-user downlink. Additional settings are summarized in Table \ref{channel_para}.

\begin{table}[h!]
\centering
\caption{Partial parameter settings.}
\label{channel_para}
\resizebox{\columnwidth}{!}{
\begin{tabular}{@{}ll@{}}
\toprule
\textbf{Parameter} & \textbf{Value} \\ 
\midrule
Noise power density $N_{0}$ & -174 dBm/Hz \cite{10155303} \\ 
The highest EIRP density $\text{PSD}_{\text{EIRP}}$ & -16.82 dBW/4 kHz \cite{fcc2020spacex}\\
Center frequency $f_{c}$ & 11.7 GHz \cite{fcc2020spacex} \\ 
Bandwidth of each LEO satellite link $B$ & 20 MHz\cite{10848151}\\
Rician factor of RU link $K^{RU}$ & 10dB \cite{10848151} \\
Rician factor of LR link $K^{LR}$ & 15dB \cite{10848151}
\\
Maximum UAV flying speed $v_{n}^{\max}$ & 12 m/s \cite{10771971}\\
Angular velocity of each LEO satellite $\omega$ & 0.001076 rad/s \cite{fcc2020spacex}
\\
\bottomrule
\end{tabular}}
\end{table}

The simulation is conducted on the Gadi supercomputer of The National Computational Infrastructure (NCI) Australia, coding by Python 3.9.2 and PyTorch 1.10.0. The setup consists of 5 local nodes and 1 master node. To improve training efficiency, the local nodes run in parallel on 12 CPU cores, while the master node operates on a single GPU.

The batch sizes for local and global training are set to $B \in [60, 70]$ and $b = 32$, respectively. A total of $3 \times 10^4$ experience traces are collected for training. The local and master nodes adopt the same policy network architecture, consisting of three hidden layers with 1024, 512 and 256 units. The hidden layers use Tanh activation, while the output activation is Sigmoid.

The output layer contains two branches: continuous outputs for UAV flight control and discrete outputs for RIS phase control, forming 2 Beta distribution (2 actions, each parameterized by 2 parameters), and the 120 Categorical distribution (50 categories for each of the 120 RIS elements). In total, the policy samples $(2 + M)$ actions per decision step.

To evaluate the effectiveness of the proposed algorithm, several baseline methods are considered for comparison:

\begin{itemize}
    \item \textbf{FedPG-NP}: Federated policy gradient without personalization, where all local nodes fully inherit the global policy parameters.

    \item \textbf{FedPG-FP}: Federated policy gradient with fixed personalization, where the partition between local and global layers remains unchanged throughout training.

    \item \textbf{SVRPG}: A non-federated baseline, in which each local node independently learns its policy using the SVRPG algorithm.
\end{itemize}

\subsection{Training Performance}\label{PerfComp}

To compare training performance, five independent runs are conducted for all methods ($[\sigma_{\text{close}},\sigma_{\text{far}}]=[2.5,3.0]$, $e_{0}=1$). 

\begin{figure}[h]
\centering
    \includegraphics[width=0.8\linewidth]{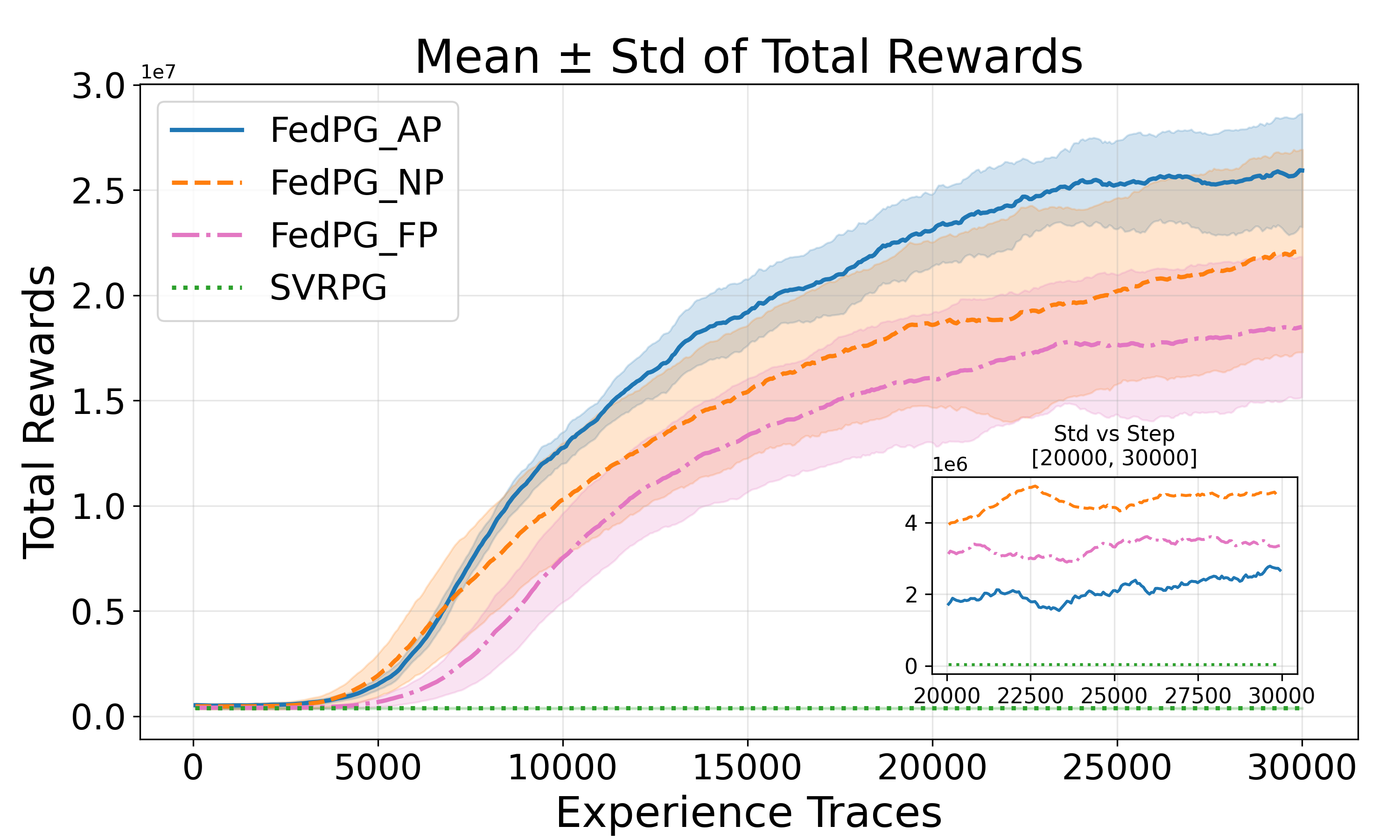}
    \caption{Training comparison of 5 runs.}
    \label{TC_reward_mean_std}
\end{figure}

Fig. \ref{TC_reward_mean_std} illustrates the mean and standard deviation (std) of the total reward over 5 independent runs, with an inset showing the std during the final 10,000 experience traces. As defined in Subsection \ref{MGF}, the total reward is defined as the sum of the average rates of active users over the period $T$. FedPG-AP achieves the highest reward and the smallest variance, indicating a stable convergence behavior. 

FedPG-NP has the largest variance, indicating strong sensitivity to environmental differences in the absence of personalization. FedPG-FP exhibits lower variance than FedPG-NP, yet its overall performance is worse, suggesting that fixed, non-adaptive personalization can be less effective than no personalization at all. SVRPG shows almost no reward improvement, implying that without federated coordination, local nodes fail to learn effectively and that most policy learning is driven by the master node.

\begin{figure}[h]
  \centering
  \begin{minipage}{\linewidth}
    \centering
    \includegraphics[width=0.8\linewidth]{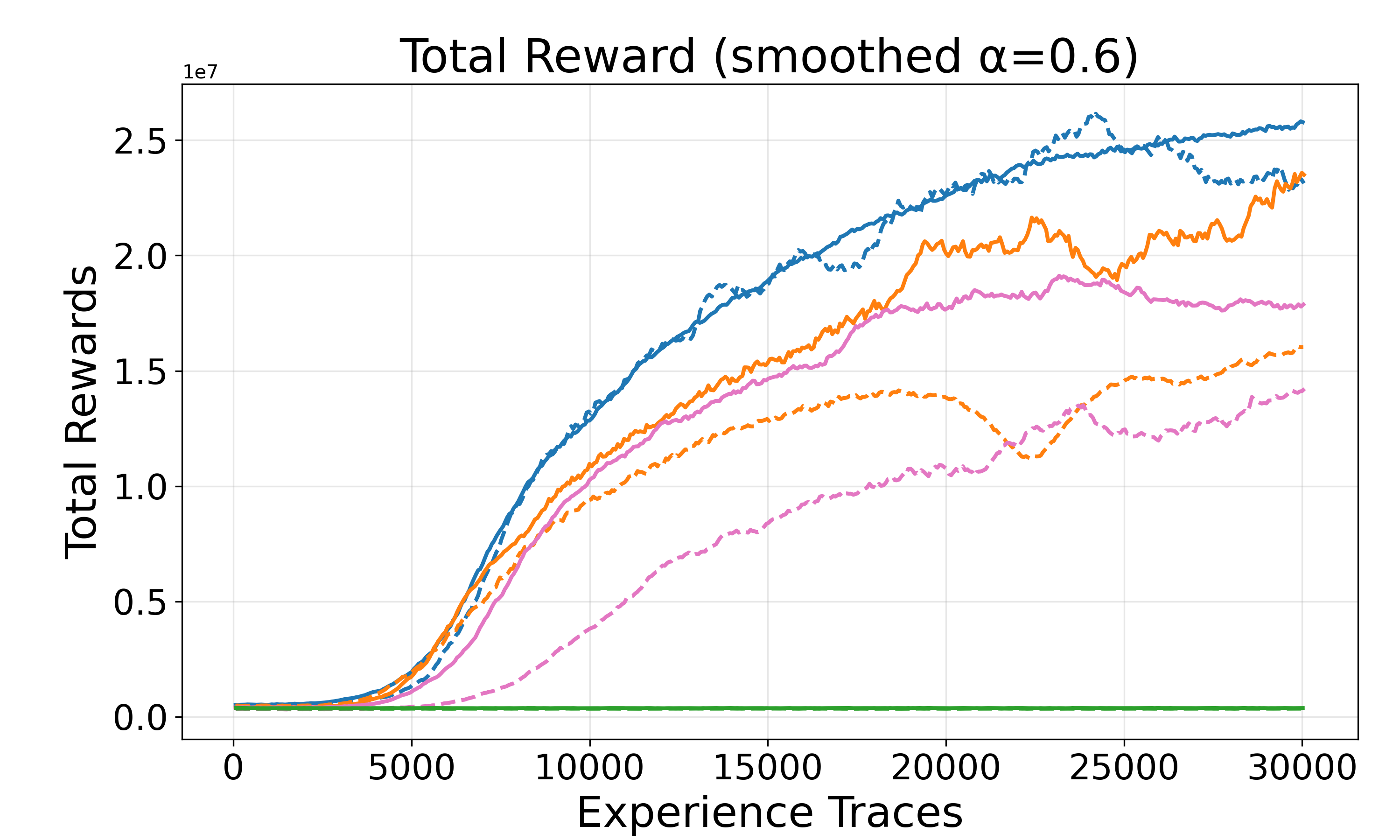}
  \end{minipage}

  \begin{minipage}{\linewidth}
    \centering
    \includegraphics[width=\linewidth]{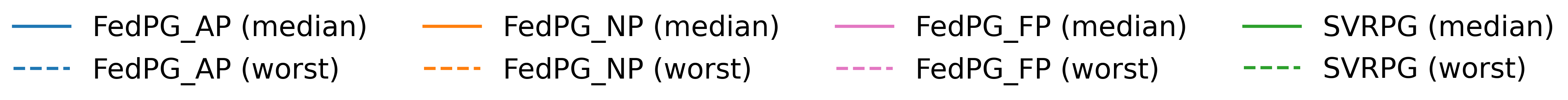}
  \end{minipage}

  \caption{Total reward comparison (median vs. worst runs).}
  \label{TC_MM_reward}
\end{figure}

To provide a finer comparison, the median and worst runs are examined. As shown in Fig. \ref{TC_MM_reward}, the worst run of FedPG-AP closely follows the median throughout training, indicating stable learning behavior. In contrast, FedPG-NP exhibits clear instability, with the worst run suffering a sharp degradation between 20,000 and 25,000 traces. For FedPG-FP, the median reward grows similarly to FedPG-NP in the early and mid stages, but slower initial learning limits its final performance, despite a steady improvement in the worst run.

\begin{figure}[h]
  \centering
  \begin{minipage}{\linewidth}
    \centering
    \includegraphics[width=0.8\linewidth]{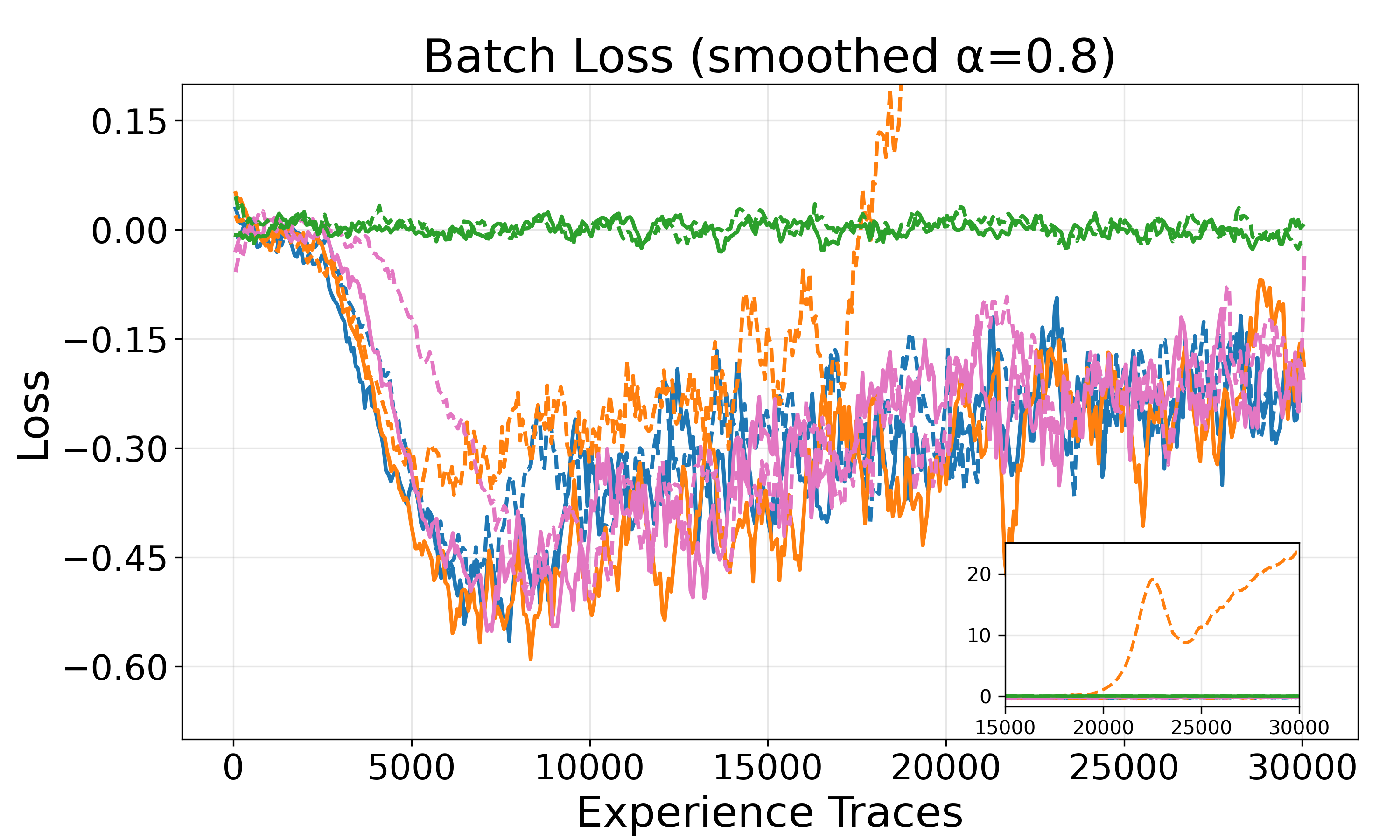}
  \end{minipage}

  \begin{minipage}{\linewidth}
    \centering
    \includegraphics[width=\linewidth]{figure/simu_result/legend_TC_MM.png}
  \end{minipage}

  \caption{Loss comparison (median vs. worst runs).}
  \label{TC_MM_loss}
\end{figure}

Fig. \ref{TC_MM_loss} presents the loss comparison. Excluding SVRPG, all methods achieve similar median performance. For FedPG-AP, the worst run is still close to the median, indicating stable performance. The worst run of FedPG-NP shows an abnormal surge, which is illustrated fully in the inset figure, suggesting convergence to a poor local optimum and explaining the reward dip. This degradation further highlights the inefficiency caused by the lack of adaptive personalization. SVRPG remains near zero throughout training, confirming its failure to learn.

\begin{figure}[h]
  \centering
  \begin{minipage}{\linewidth}
    \centering
    \includegraphics[width=0.6\linewidth]{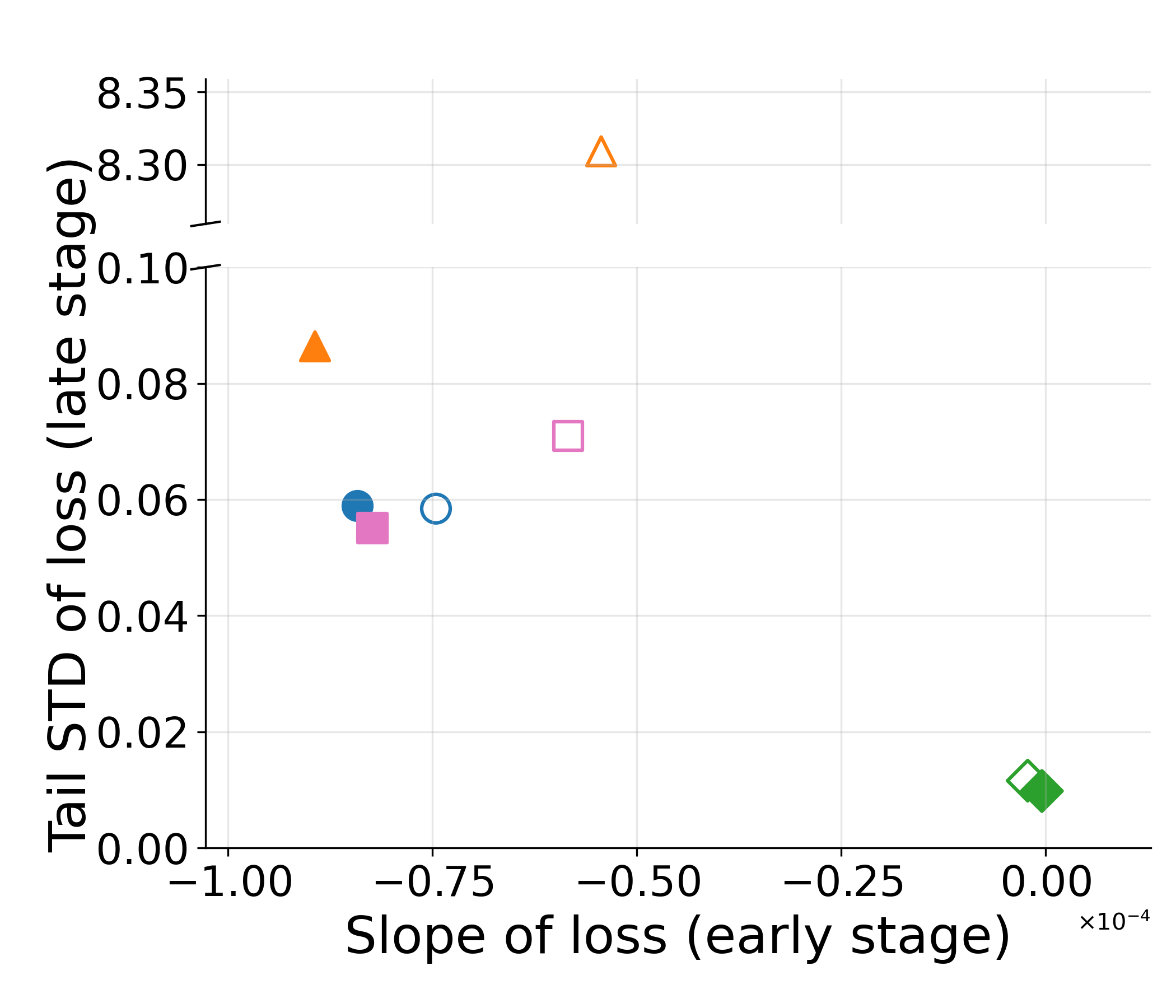}
  \end{minipage}

  \begin{minipage}{\linewidth}
    \centering
    \includegraphics[width=\linewidth]{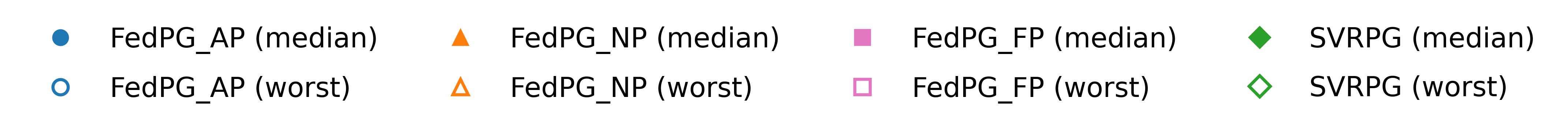}
  \end{minipage}

  \caption{Learning speed and convergence comparison (median vs. worst runs).}
  \label{TC_MM_loss_scatter}
\end{figure}

Two metrics are used to further analyze the loss behavior: the slope of a linear fit in the early stage (0–8000 traces), indicating initial learning speed, and the std over the final 10,000 traces, reflecting convergence stability. Smaller values of both metrics imply faster learning and more stable convergence. Therefore, points closer to the lower-left corner in the scatter plot shown in Fig. \ref{TC_MM_loss_scatter} indicate better performance. FedPG-AP performs well in both the median and worst runs, achieving fast learning and stable convergence. FedPG-FP shows strong median performance but a larger gap to its worst run. FedPG-NP attains the fastest median learning speed but suffers from poor stability. SVRPG remains near zero on both metrics, indicating that local policies stay close to their initial states.

These results indicate that lack of personalization causes severe instability under heterogeneous environments, whereas fixed personalization slows early learning. We therefore proceed with a finer parameter analysis to clarify how personalization influences training efficiency.

\subsection{Parameter Analysis}

A large number of parameters are involved in the training of FedPG-AP. The experimental results in Subsection \ref{PerfComp} suggest that the algorithm’s performance is particularly sensitive to personalization-related parameters. Accordingly, this subsection focuses on a detailed analysis of these parameters to examine the behavior of the personalization mechanism. In FedPG-AP, the level of personalization is primarily determined by the thresholds $\sigma_{\text{close}}$, $\sigma_{\text{far}}$, and the initial partition $e_{0}$ between local and global. 

\begin{table}[h]
\centering
\caption{Threshold pair and buffer area.}
\label{tab:dual_threshold}
\resizebox{\columnwidth}{!}{
\begin{tabular}{c|c c c c c}
\hline
$\left[\sigma_{\text{close}},\,\sigma_{\text{far}}\right]$
& $(2.0,\,2.5)$
& $[2.0,\,3.0)$
& $[2.5,\,3.0]$
& $[2.5,\,3.5]$
& $[3.0,\,3.5]$ \\
\hline
Buffer area $A$
& $2.25\pi$
& $5.00\pi$
& $2.75\pi$
& $6.00\pi$
& $3.25\pi$ \\
\hline
\end{tabular}}
\end{table}

Taking the median gradient as the center, the two thresholds define an annular region, referred to as the buffer region. Table \ref{tab:dual_threshold} lists different threshold pairs in ascending order and the corresponding buffer area, computed as $A = \pi(\sigma_{\text{far}}^2 - \sigma_{\text{close}}^2)$.

\begin{figure}[h]
\centering
    \includegraphics[width=0.8\linewidth]{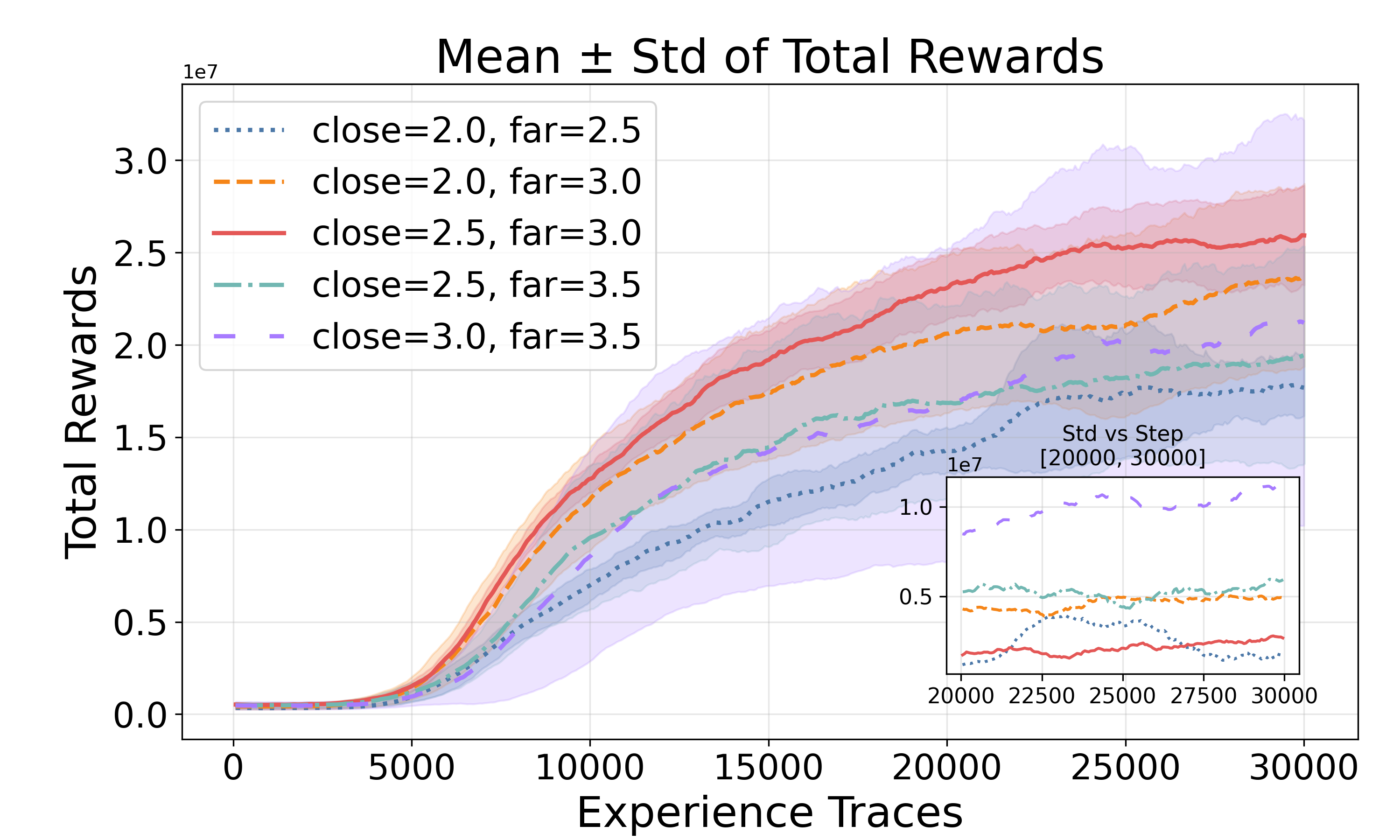}
    \caption{Threshold comparison of 5 runs.}
    \label{PA_thre_reward_mean_std}
\end{figure}

Five independent runs are conducted for each threshold pair ($e_{0} =1 $). Fig. \ref{PA_thre_reward_mean_std} shows the mean and standard deviation of the total rewards across all runs, with an inset illustrating the late-stage variance. $(2.5,3.0)$ achieves the best trade-off, attaining the highest reward with the smallest variance. Although $(2.0,3.0)$ and $(2.5,3.5)$ are close to $(2.5,3.0)$ in distance settings, their larger buffer areas make the behavior closer to FedPG-FP, improving stability but reducing reward. 

By contrast, $(2.0, 2.5)$ and $(3.0, 3.5)$ yield buffer regions comparable to that of $(2.5, 3.0)$ but are overly extreme. $(2.0, 2.5)$ is overly tight, as it triggers global enhancement prematurely and then quickly switches to personalization, which restricts learning and leads to a lower reward. $(3.0, 3.5)$ is overly loose, as excessive early personalization increases sensitivity to heterogeneous environments and results in substantially higher variance, performing similar to FedPG-NP.

\begin{figure}[h]
  \centering
  \begin{minipage}{\linewidth}
    \centering
    \includegraphics[width=0.8\linewidth]{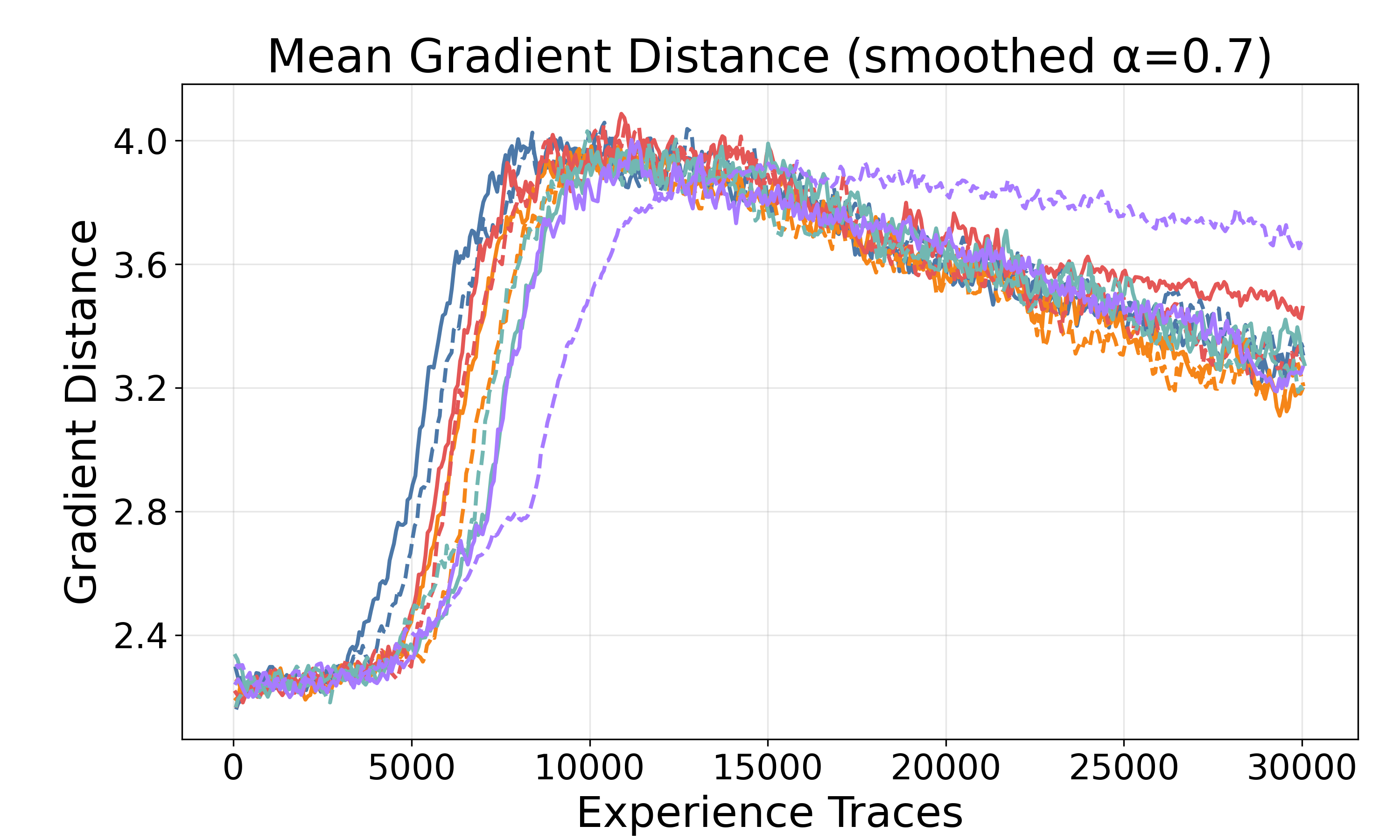}
  \end{minipage}

  \begin{minipage}{\linewidth}
    \centering
    \includegraphics[width=\linewidth]{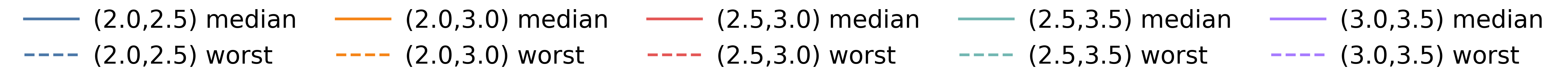}
  \end{minipage}

  \caption{Gradient distance comparison under different threshold pairs (median vs. worst runs).}
  \label{PA_thre_grad}
\end{figure}

As analyzed in Subsection \ref{AP}, the gradient distance reflects the degree of policy divergence. To provide a finer-grained comparison among different threshold pairs, Fig. \ref{PA_thre_grad} illustrates the evolution of the mean distance between local gradients for the median and worst runs. 

$(2.0,2.5)$ triggers global enhancement earliest, which induces rapid policy divergence and the fastest rise in gradient distance, aligning with the fast initial learning of FedPG-NP. The $(2.5,3.0)$ pair achieves a balanced divergence rate with closely aligned median and worst runs. Larger buffer areas in $(2.0,3.0)$ and $(2.5,3.5)$ improve stability but slow policy divergence, while the overly loose $(3.0,3.5)$ delays global enhancement and results in insufficient early divergence.

\begin{figure}[ht]
\centering
    \includegraphics[width=0.8\linewidth]{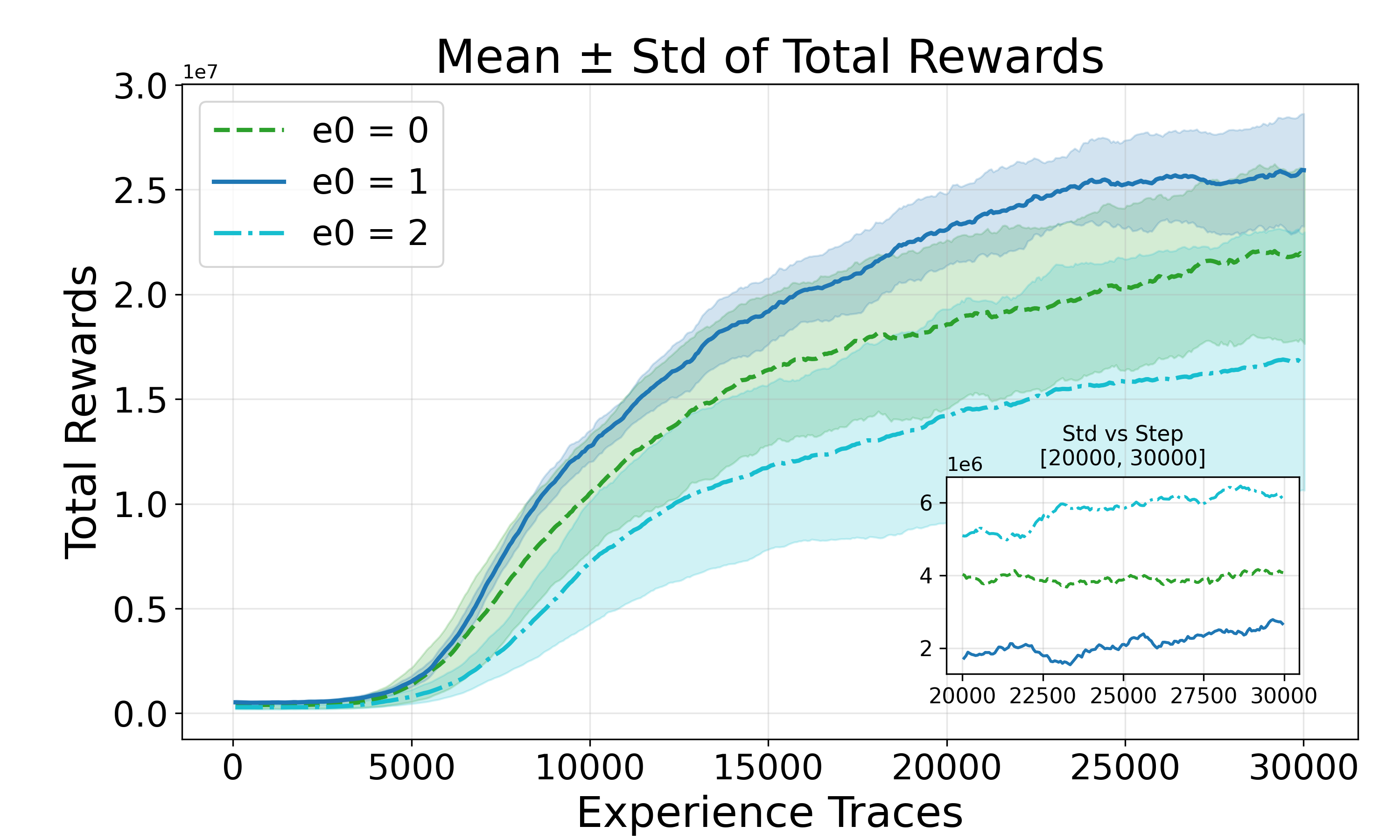}
    \caption{Initial partition comparison of 5 runs.}
    \label{PA_enh_reward_mean_std}
\end{figure}

\begin{figure}[h]
  \centering
  \begin{minipage}{\linewidth}
    \centering
    \includegraphics[width=0.8\linewidth]{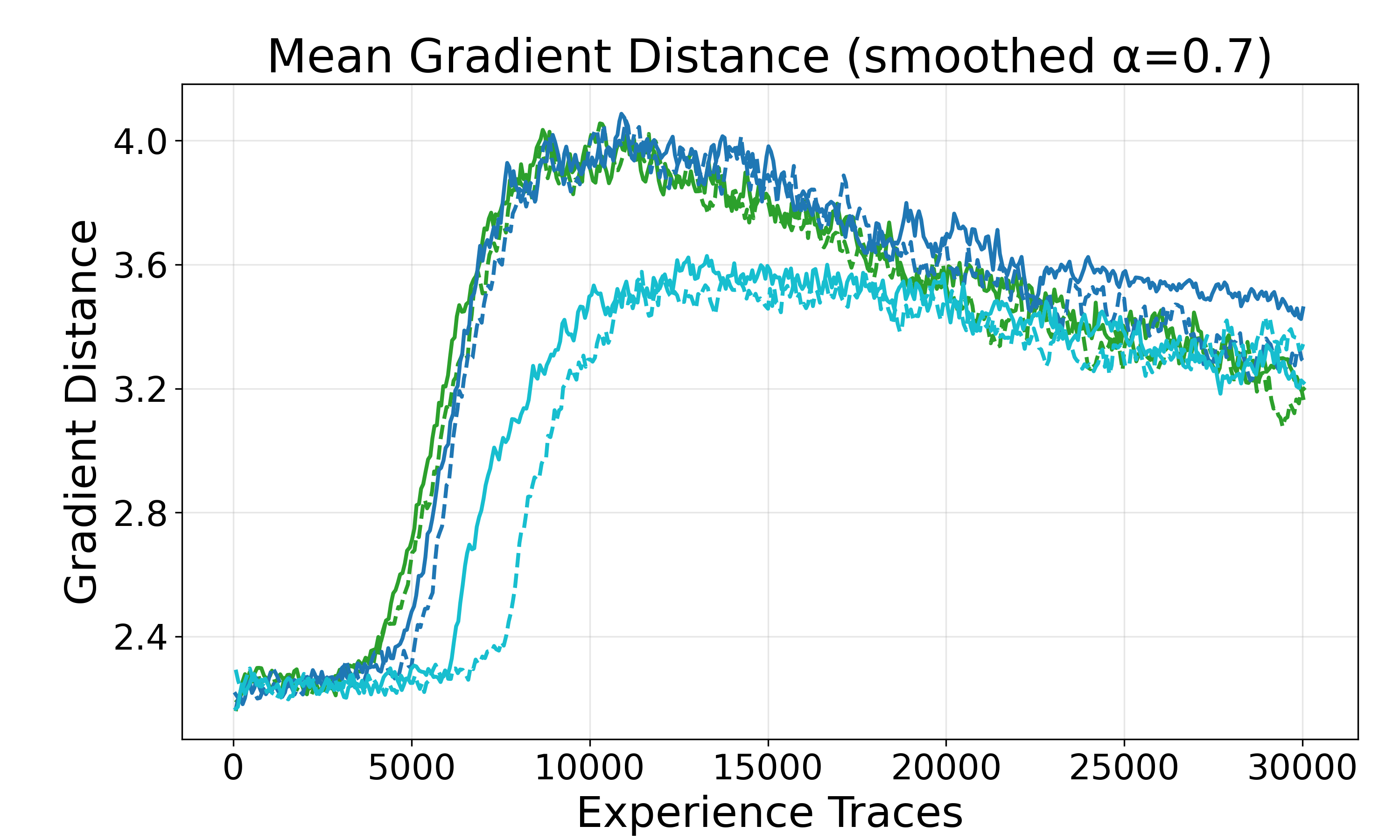}
  \end{minipage}

  \begin{minipage}{\linewidth}
    \centering
    \includegraphics[width=\linewidth]{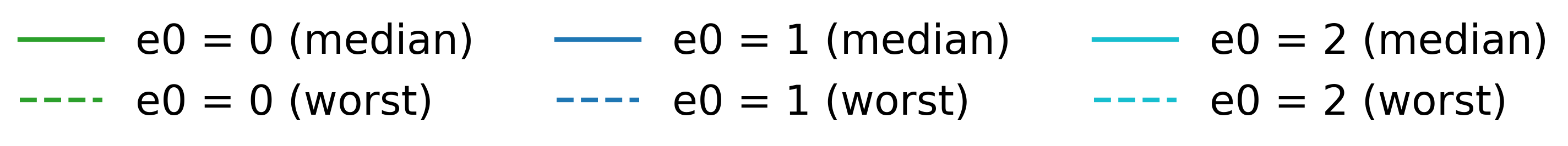}
  \end{minipage}

  \caption{Gradient distance comparison under different initial partition (median vs. worst runs).}
  \label{PA_enh_grad}
\end{figure}

The initial partition $e_{0}$ fundamentally determines how parameters are inherited across layers at the beginning of each epoch. Since the policy network consists of three hidden layers as set, $e_{0} \in \{0,1,2\}$ ($(\sigma_{\text{close}}$, $\sigma_{\text{far}}) = (2.5, 3.0)$). Similarly, Fig. \ref{PA_enh_reward_mean_std} illustrates the mean ± std of total rewards over five runs for different values of $e_{0}$. Fig. \ref{PA_enh_grad} further reports the mean gradient distance for different $e_{0}$, evaluated on the corresponding median and worst runs.

By jointly examining Figs. \ref{PA_enh_reward_mean_std} and \ref{PA_enh_grad}, $e_{0} = 1$ maintains a balanced interaction between local and global layers, achieving the best overall performance with the highest reward and the lowest variance. When $e_{0} = 0$, all layers default to global parameters unless local enhancement is triggered; although policy divergence occurs most rapidly, the resulting reward performance remains suboptimal. In contrast, when $e_{0} = 2$, only the last hidden layer remains global. This overly strong personalization promotes early exploration but leads to insufficient learning, resulting in the poorest performance, characterized by the lowest reward, the largest variance, and slow, limited policy divergence.

Taken together, the threshold and partition studies point to a consistent conclusion: a balanced configuration between personalization and global sharing offers the best performance. Such balanced settings provide sufficient room for exploration through personalization, while maintaining the stability and generality offered by global sharing, thereby enabling effective adaptation and performance increases in heterogeneous environments. These results further underscore the advantage of FedPG-AP over both FedPG-FP and FedPG-NP.

\subsection{Validation}

To evaluate the effectiveness of the trained policies, the median performance policies are used for each algorithm. Each policy is tested for 100 independent runs, where the UAV executes its flight-control actions and RIS phase-control actions to determine the mean downlink rate of active users at each time $t$ over the full period $T$. Each run is conducted in a different randomly generated environment. The results of the 100 runs are summarized as follows.

\begin{figure}[h]
\centering
\begin{minipage}[t]{0.49\linewidth}
    \centering
    \includegraphics[width=\linewidth]{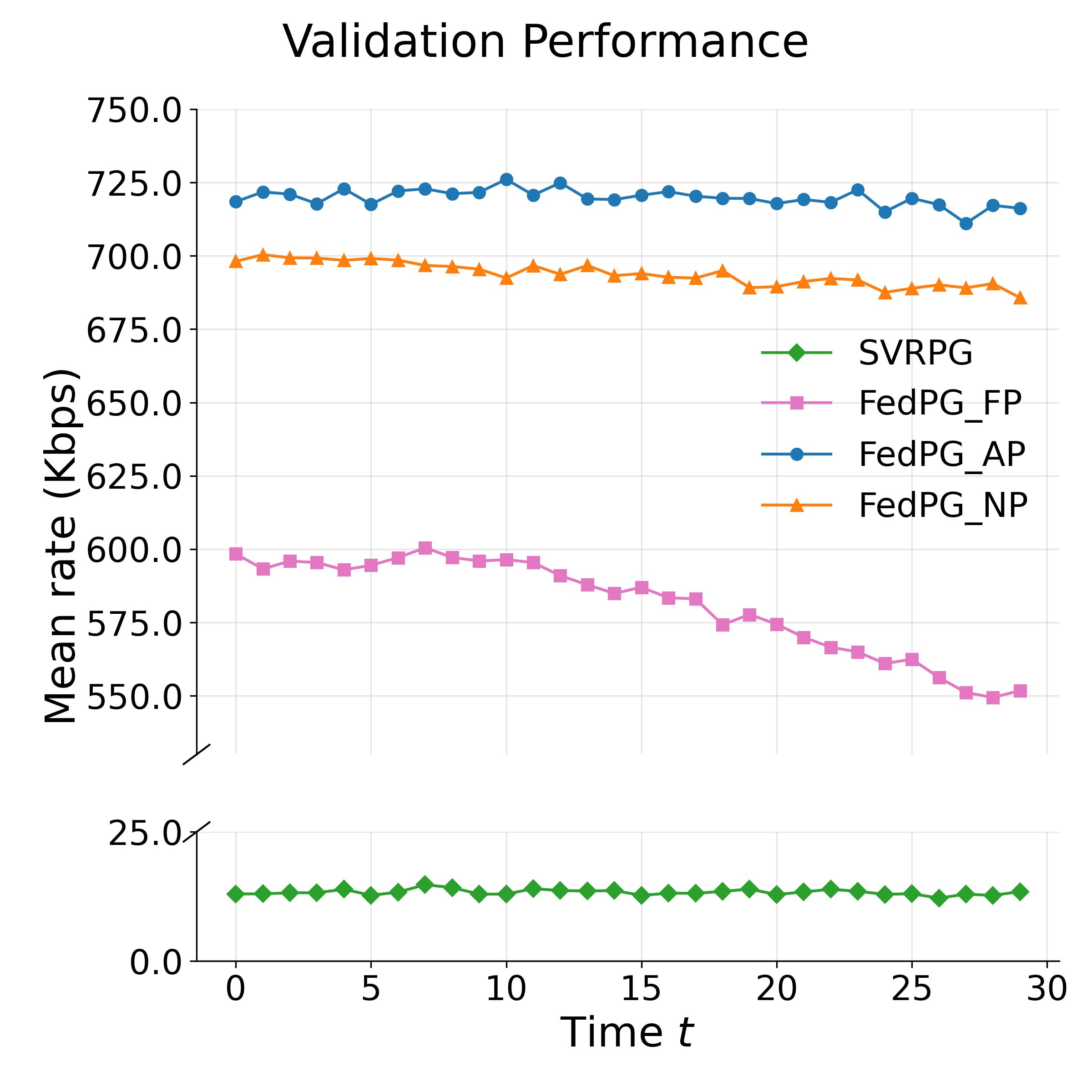}
    \caption{Comparison of downlink rate.}
    \label{eval_rate}
\end{minipage}
\hfill
\begin{minipage}[t]{0.49\linewidth}
    \centering
    \includegraphics[width=\linewidth]{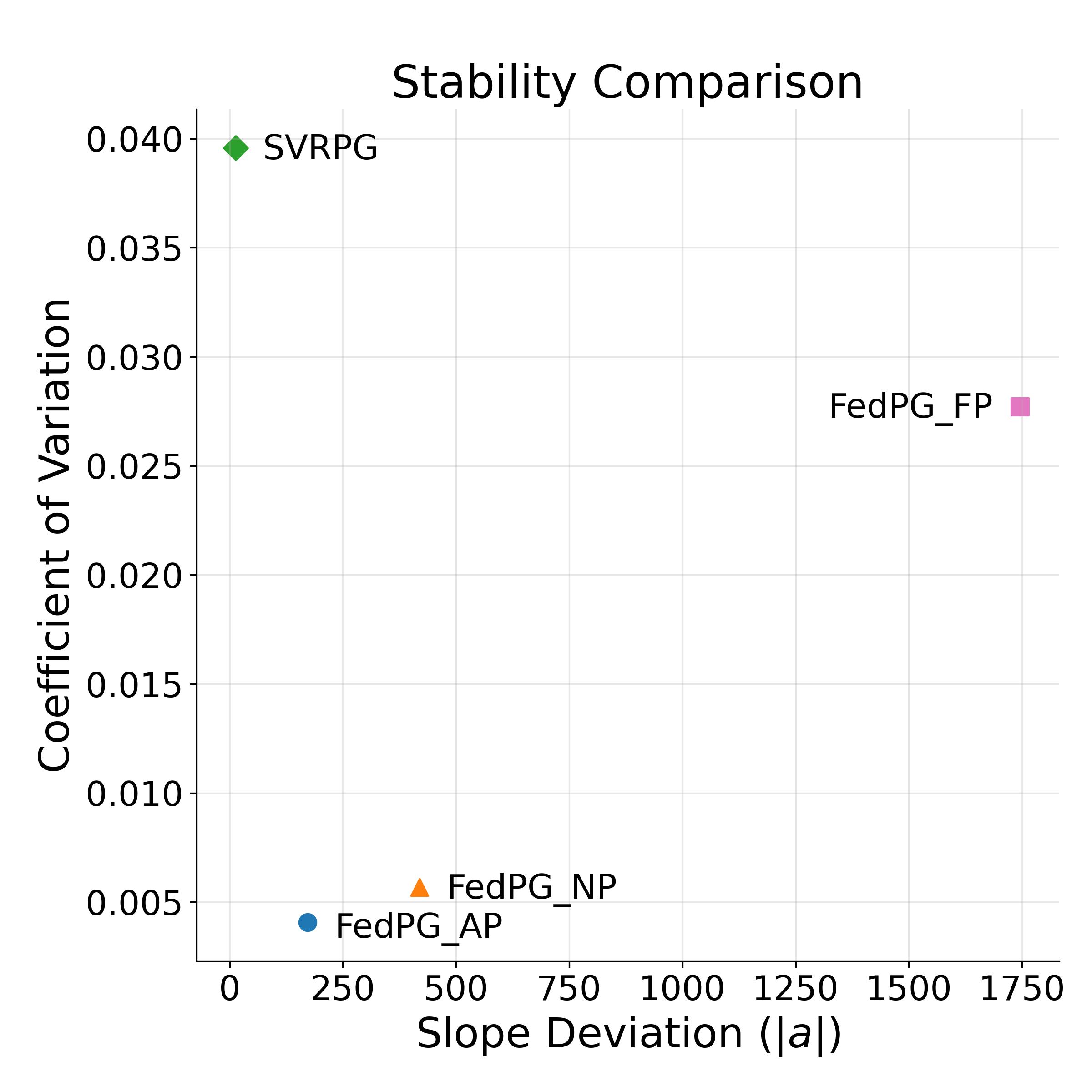}
    \caption{Stability of downlink rate.}
    \label{eval_cv_slope_scatter}
\end{minipage}
\end{figure}

Fig. \ref{eval_rate} shows the average mean downlink rate over time across 100 runs. The proposed FedPG-AP method consistently maintains the highest rate throughout the entire period, reaching approximately 725 Kbps. This is followed by FedPG-NP and FedPG-FP, while SVRPG performs the worst and is even an order of magnitude lower than the other methods. These results are fully consistent with the training results.

For further evaluation, two metrics are considered. The coefficient of variation (CV), defined as the ratio of the std to the mean rate over 100 runs, quantifies the relative fluctuation of the rate, where smaller values indicate more stable and reliable transmission. The slope deviation (SD) is defined as the absolute slope obtained from a linear fit of the mean rate over time. Since discounted reward training emphasizes early-stage actions, which may lead to gradual performance degradation, SD captures the severity of this degradation. As shown in Fig. \ref{eval_cv_slope_scatter}, the proposed FedPG-AP achieves the best overall performance, exhibiting the smallest CV (most stable transmission) and the smallest SD (minimal degradation). FedPG-NP ranks second but suffers from a more noticeable decline. FedPG-FP shows clear fluctuations and degradation over time. Although SVRPG has the smallest SD, this result is not informative given its extremely low rate.

Each simulation run is conducted under a randomly generated environment, the stability of the downlink rate validates the strong adaptability of the proposed FedPG-AP to diverse environments. Therefore, we do not separately retrain and validate the model under different system-level parameters, such as the number of hotspots $N$, the number of RIS elements $M$, and the number of ports $H$. Instead, we aim to provide a practical personalized FRL framework for handling environmental heterogeneity. A more comprehensive investigation of system-level parameter variations is left for future work.

\section{Conclusion}\label{conclusion}

This paper has investigated a comprehensive SAGIN integrating RIS and FAS, and has developed a complete channel model from a LEO satellite constellation to ground users via RIS-assisted UAV relays, explicitly accounting for both FAS-equipped and conventional users. A system-level downlink rate maximization problem has been formulated and analyzed via a hierarchical Stackelberg game. Then, a personalized FRL algorithm has been proposed to jointly optimize UAV trajectories and RIS phase controls under heterogeneous environments.

Simulation results have demonstrated that personalization is critical to improving FRL performance in heterogeneous SAGINs, and adaptability is necessary to balance local specialization and global knowledge sharing. Future work will further explore the impact of system heterogeneity and scalability on learning behavior, aiming to provide deeper insights into the design of large-scale SAGINs.

\bibliographystyle{IEEEtran}
    \bibliography{mylib}

\end{document}